\documentclass[%
 reprint,
 superscriptaddress,
 longbibliography,
 preprintnumbers,
 amsmath,amssymb,
aps,
 prb,
floatfix,
]{revtex4-2}

\usepackage{graphicx}
\usepackage{dcolumn}
\usepackage{bm}
\usepackage{hyperref}
\usepackage{xcolor}
\usepackage{braket}
\usepackage{mathtools}
\usepackage{subfigure}

\usepackage{physics}
\usepackage{siunitx}
\usepackage[mode=buildnew]{standalone}

\usepackage[maxfloats=256]{morefloats}
\begin{document}

\preprint{}

\title{Theory of infrared double-resonance Raman spectrum in graphene: the role of the zone-boundary electron-phonon enhancement}
\author{Lorenzo Graziotto}
\affiliation{Department of Physics, Sapienza University of Rome, Piazzale Aldo Moro 5, 00185 Rome, Italy}
\affiliation{Institute for Quantum Electronics, ETH Zürich, Auguste-Piccard-Hof 1, 8093 Zürich, Switzerland\looseness=-1}
\author{Francesco Macheda}
\affiliation{Department of Physics, Sapienza University of Rome, Piazzale Aldo Moro 5, 00185 Rome, Italy}
\affiliation{Istituto Italiano di Tecnologia, Graphene Labs, Via Morego 30, 16163 Genova, Italy}
\author{Thibault Sohier}
\affiliation{Laboratoire Charles Coulomb (L2C), Université de Montpellier, CNRS, 34095 Montpellier, France}
\author{Matteo Calandra}
\affiliation{Department of Physics, University of Trento, Via Sommarive 14, 38123 Povo, Italy}%
\author{Francesco Mauri}
\affiliation{Department of Physics, Sapienza University of Rome, Piazzale Aldo Moro 5, 00185 Rome, Italy}
\affiliation{Istituto Italiano di Tecnologia, Graphene Labs, Via Morego 30, 16163 Genova, Italy}%
\begin{abstract}
We theoretically investigate the double-resonance Raman spectrum of monolayer graphene down to infrared laser excitation energies. By using first-principles density functional theory calculations, we improve upon previous theoretical predictions based on conical models or tight-binding approximations, and rigorously justify the evaluation of the electron-phonon enhancement found in Ref.~[Venanzi, T., Graziotto, L.\ \emph{et al.}, Phys.\ Rev.\ Lett.\ \textbf{130}, 256901 (2023)]. We proceed to discuss the effects of such enhancement on the room temperature graphene resistivity, hinting towards a possible reconciliation of theoretical and experimental discrepancies.
\end{abstract}

\maketitle


\section{Introduction}
Raman spectroscopy is a widespread and versatile experimental technique used to characterize graphitic materials. In particular, in few layer graphene it is commonly used to determine the number of layers~\cite{PhysRevLett.97.187401,Gupta2006,ferrari2007raman,herziger2014two,Graf2007}, carrier and defect concentrations~\cite{Beams_2015,PhysRevLett.97.266407,PhysRevLett.98.166802,Eckmann2012},  as well as phonon properties~\cite{PhysRevB.76.233407,Berciaud2013,graziotto2022probing}.

Raman scattering refers to the inelastic scattering of light by a molecular or crystalline sample, due to the concurrent excitation (Stokes processes) or de-excitation (anti-Stokes processes) of vibrational degrees of freedom of the sample. In a single-particle description, in the Stokes case, the incoming photon (of frequency $\omega_L$) creates $n$ phonons with total energy $\hbar \Omega_\text{ph}^\text{tot}$ and then leaves the sample with a frequency $\omega_L - \Omega_\text{ph}^\text{tot}$. In defect-free monolayer graphene, for laser excitation energies $\epsilon_L = \hbar \omega_L$ up to the near-UV i.e. with wavevectors much smaller than the size of the first Brillouin zone (FBZ), conservation of crystal momentum requires the sum of the wavevectors of the Raman scattered phonons to be equal to zero. The one-phonon Stokes spectrum of defect-free monolayer graphene (simply referred to as graphene in the following) consists of the so-called G peak only, which is due to the excitation of a single phonon at the $\bm{\Gamma}$ point of the FBZ. The two-phonon spectrum is instead explained within the double-resonance scheme~\cite{venezuela2011theory}, which amounts to considering the intermediate role played by electrons and holes in the creation of the phonons pair. Indeed, the sharpness and separation of the two-phonon lines arise due to the condition that the intermediate states of the Raman process can be eigenstates of the system, with a lifetime given by their many-body interaction. The double-resonance scheme is then depicted as an excitation by the incoming photon of an electron-hole pair which excites a phonon through the electron-phonon coupling (EPC), and then recombines with the emission of the outgoing photon. Conservation of crystal momentum requires that the two phonons, which belong to either the same or to different optical branches, have opposite momenta. The most relevant double-resonance peaks are the so-called 2D and 2D': the former relates to the case in which both the electron and the hole scatter between two different Dirac cones, hence the pair of phonons has (opposite) wavevectors which are close to the edge of the FBZ (\textbf{K} and \textbf{K'} points in reciprocal space); the latter relates to the case in which the scattering happens within the same Dirac cone, hence the pair of phonons has wavevectors close to the center of the FBZ ($\bm{\Gamma}$ point in reciprocal space).

As addressed below, the intensity of the 2D (2D') peak scales with the fourth power of the EPC evaluated at the \textbf{K} ($\bm{\Gamma}$) point, hence the ratio of the two gives a clear indication of how the EPC at \textbf{K} evolves with respect to the EPC at $\bm{\Gamma}$ as a function of the excitation energy. In particular, Raman measurements performed at a laser energy of \SI{1.16}{eV} in  Ref.~\cite{graziotto2022probing} show that the EPC is enhanced at zone-boundary while approaching the Dirac cone. This is due to an underlying enhancement of the electron-electron interaction, firstly predicted in Refs.~\cite{basko2008interplay, lazzeri2008impact}, which can be taken into account via an excitation energy dependent coefficient. In this work, we rigorously justify and test the boundaries of the theoretical analysis of Ref.~\cite{graziotto2022probing} using density functional theory (DFT) \emph{ab initio} calculations. As opposed to previous works which employed the tight-binding approximation~\cite{venezuela2011theory}, our approach allows us to convincingly put the analytical results of Ref.~\cite{basko2008theory} to the test, in particular regarding the role of the electron-hole asymmetry, of the trigonal warping of both phononic and electronic dispersions, and of the inverse lifetime of the electronic states in determining the line-shape and the integrated area of the Raman peaks.

The present results are also used to shed light on a open question regarding the electronic transport of graphene. Ref.~\cite{PhysRevB.90.125414} and \cite{park2014electron} showed that the \emph{ab initio} resistivity of graphene computed via Boltzmann transport equation largely underestimates the experimental value in the equipartition regime ($T>\SI{270}{K}$) especially at low doping levels.
In those works, it was argued that this could come from the enhancement of the zone-boundary EPC studied here rather than other extrinsic mechanisms like remote polar-optical phonons from the substrate. 
However, the enhancement needed to explain the resistivity measurements appeared quite large at the time. We discuss how the values obtained via Raman in Ref.~\cite{graziotto2022probing} are consistent with such a large enhancement, although they are not directly applicable to transport due to different doping setups.

The paper is organized as follows: in Section~\ref{sec:theo_frame} we introduce the model Hamiltonian and the formalism to calculate the Raman scattering intensity (showing also a simplified model to clarify the double-resonance mechanism). In Sec.~\ref{sec:Comp_Appr} we present the computational details of the implementation of the calculation. In Sec.~\ref{sec:Res_Disc} we show the results of the calculation of the Raman spectrum, and its dependence on the various parameters involved, and discuss its physical implications, in particular its impact on the room temperature resistivity. Finally, Sec.~\ref{sec:Conc} summarizes the main findings and outlines the possibilities for further investigations. 

\section{Theoretical framework}\label{sec:theo_frame}

\label{subsec:DFTHamiltonian}
We treat electrons, phonons, the EPC and the response to external perturbations within DFT. The DFT ground-state $\ket{\mathrm{GS}}$ corresponds to the Fermi sea, i.e.\ all the electronic states below the Fermi energy $\epsilon_F$ are occupied, and all the states above $\epsilon_F$ are empty. In graphene $\epsilon_F$ corresponds to the energy of the electronic state having wavevector \textbf{K}, for which the conduction and the valence bands are degenerate. Introducing fermionic creation and annihilation operators $c_{\vb{k}}^{\alpha\dagger}, c_{\vb{k}}^\alpha$ for an electron belonging to band $\alpha$, with wavevector $\vb{k}$ and energy  $\epsilon_{\vb{k}}^\alpha$, we may write
\begin{equation}
    \ket{\mathrm{GS}} = \prod_{\vb{k}v} c_{\vb{k}}^{v\dagger} \ket{0}_{\text{el}}\otimes\ket{0}_{\text{phot}}\otimes\ket{0}_{\text{ph}},
\end{equation}
where $\ket{0}_{\text{el}/\text{phot}/\text{ph}}$ indicates the non-interacting electronic/photonic/phononic vacuum and $v$ indicates valence states; we also call $\ket{\mathrm{eGS}} = \prod_{\vb{k}v} c_{\vb{k}}^{v\dagger} \ket{0}_{\text{el}}$ and omit $\otimes$ and the subscript ${}_\text{el/phot/ph}$ in the following, if not strictly necessary.
The electronic Hamiltonian is given by
\begin{equation}
    \mathcal{H}_\mathrm{KS} =\frac{\mathbf{p}^2}{2m_e}+V_{\textrm{KS}}(\mathbf{r})=\sum_{\vb{k},\alpha} \epsilon_{\vb{k}}^\alpha c_{\vb{k}}^{\alpha\dagger} c_{\vb{k}}^\alpha,
\end{equation}
where $m_e$ is the electronic mass. Since the Kohn-Sham band structure is only an approximate description of the true excitation spectrum, the quasiparticles of the system will have complex eigenvalues with non-zero imaginary part, as it will be more carefully addressed below. The interaction of the electrons with the electromagnetic field is included in the Hamiltonian via the minimal coupling, discarding the $\mathbf{A}^2$ term since we are interested only in resonant processes:
\begin{equation}
\mathcal{H}_\mathrm{el-em} = - \frac{e}{2m_e}\sum_i\left[\vb{p}_i\cdot \vb{A}(\vb{r}_i)+\vb{A}(\vb{r}_i)\cdot \vb{p}_i  \right].
\end{equation}

It is convenient to quantize the electromagnetic field, introducing bosonic creation and annihilation operators $a^{\sigma\dagger}_{\vb{p}}$, $a^{\sigma}_{\vb{p}}$ for a photon with polarization $\sigma$, wavevector $\vb{p}$, and energy $\hbar \omega_{\vb{p}} = \hbar c p$ where $c$ is the speed of light and $p \equiv \abs{\vb{p}}$. The second-quantized vector potential~\cite{cohen1989photons} is given in the Coulomb gauge and in the interaction picture by 
\begin{equation}
\begin{split}
\vb{A}(\vb{r}, t) = \sum_{\vb{p},\sigma} \sqrt{\frac{\hbar}{2\varepsilon_0 V \omega_{\vb{p}}}} \big(\hat{\epsilon}_{\vb{p}}^{\sigma} a_{\vb{p}}^{\sigma}(t) e^{i \vb{p}\vb{r}} + \\    
+ \hat{\epsilon}^{\sigma*}_{\vb{p}} a^{\sigma\dagger}_{\vb{p}}(t) e^{-i \vb{p}\vb{r}} \big),
\end{split}
\end{equation}
where $V$ indicates the volume inside which the field is quantized, $\hat{\epsilon}_{\vb{p}}^{\sigma}$ is the polarization vector for a photon of polarization $\sigma$ and wavevector $\vb{p}$, and the time dependence of the creation and annihilation operators is given trivially by $a_{\vb{p}}^{\sigma}(t) = e^{-i \omega_{\vb{p}} t} a_{\vb{p}}^{\sigma}$, and the corresponding adjoint equation; to ease the notation, we will avoid to explicit the polarization index if not necessary. The free electromagnetic field Hamiltonian is given as usual by
\begin{equation}
    \mathcal{H}_{\mathrm{em}} = \sum_{\vb{p},\sigma}\hbar \omega_{\vb{p}} \left(a_{\vb{p}}^{\sigma\dagger }a_{\vb{p}}^{\sigma} + \frac{1}{2}\right).
\end{equation}

The phonon Hamiltonian is given by
\begin{equation}
    \mathcal{H}_\mathrm{ph} = \sum_{\vb{q},\nu} \hbar \omega_{\vb{q}}^\nu \left(b_{\vb{q}}^{\nu\dagger} b_{\vb{q}}^\nu + \frac{1}{2}\right),
\end{equation}
 where we have introduced bosonic creation and annihilation operators $b_{\vb{q}}^{\nu\dagger}, b_{\vb{q}}^\nu$ for a phonon with wavevector $\vb{q}$ belonging to branch $\nu$ with frequency $\omega_{\vb{q}}^\nu$, which is already dressed by the electron-phonon interaction, i.e.\ the only ones that will appear in the Raman diagrams; even these quasiparticles have finite lifetime that will be taken into account in the following.
The electron-phonon interaction is introduced via the following interaction Hamiltonian in the Born-Oppenheimer approximation~\cite{giustino2017electron} 
\begin{equation}
    \mathcal{H}_\mathrm{el-ph} = \frac{1}{\sqrt{N_p}}\sum_{\substack{\vb{k},\vb{q} \\ \alpha,\alpha',\nu}} g_{\vb{k}+\vb{q}\alpha,\vb{k}\alpha'}^\nu c^{\alpha\dagger}_{\vb{k+q}} c^{\alpha'}_{\vb{k}} (b_{\vb{q}}^\nu + b_{\vb{-q}}^{\nu\dagger}),
\end{equation}
where $N_p$ is the number of unit cells in the Born-Von Karman supercell and $g$ is the EPC matrix element. If we consider the atoms positioned at ${\bf u}^{\mathbf{R}}_{s}=\mathbf R+\boldsymbol{\tau}_s $, where $\mathbf{R}$ is a Bravais vector and $\boldsymbol{\tau}_s$ a basis vector, then collective displacements with wavevector $\mathbf{q}$ of the atoms $s$ along the Cartesian axis $\beta$ induce a cell-periodic potential variation $V^{\mathbf{q}}_{s\beta}(\mathbf{r})$ defined by
\begin{align}
 V^{\mathbf{q}}_{s\beta}(\mathbf{r})  =e^{-i{\mathbf q}\cdot \mathbf{r}} \pdv{V_{\text{KS}}(\mathbf r)}{  u^{\mathbf{q}}_{s\beta}} =e^{-i{\mathbf q}\cdot \mathbf{r}}\sum_{\mathbf R} e^{i{\mathbf q}\cdot( \mathbf{R} +\bm{\tau}_s  ) }  \pdv{V_{\text{KS}}(\mathbf r)}{  u^{\mathbf{R}}_{s\beta}} .
\end{align} 
Passing in the phonon eigenvector basis, the EPC matrix elements may then be written as
\begin{equation}
\label{eq:EPC}
\begin{aligned}
g_{\vb{k}+\vb{q}\alpha,\vb{k}\alpha'}^\nu &= \sqrt{\frac{\hbar}{2M\omega_{\vb{q}}^\nu}} \mel{\vb{k}+\vb{q},\alpha}{ V^{\mathbf{q}}_{\nu}(\mathbf{r})}{\vb{k},\alpha'}   \\
 &= \sqrt{\frac{\hbar}{2M\omega_{\vb{q}}^\nu}}\, D_{\vb{k}+\vb{q}\alpha,\vb{k}\alpha'}^\nu,
    \end{aligned}
\end{equation}
where $\bra{\mathbf{k},\alpha}$ (and its conjugate) are the cell-periodic parts of the Bloch wavefunctions, $M$ is the carbon atomic mass and $V^{\mathbf{q}}_{\nu}(\mathbf{r})$ is obtained by the contraction of $V^{\mathbf{q}}_{s\beta}(\mathbf{r})$ with the phonon mode of polarization $\nu$.

\subsection*{Double-resonance scattering intensity}
We will limit ourselves to the treatment of Stokes processes only, since in monolayer graphene at room temperature they are the dominant ones. Our approach consists in computing the two-phonon Raman intensity via the theory of scattering. Although following the same approach of Ref.~\cite{venezuela2011theory}, we pay particular care in this work to keep into account all the energetic factors that appear in the intensity, since we are interested in comparing the results for different excitation energies. We employ Fermi's golden rule generalized to the fourth perturbative order given that, being Raman scattering a two-photon process, the second perturbative order in the interaction of the electrons with the electromagnetic field is needed, and to deal with the two-phonon spectrum we need two more perturbative orders in the electron-phonon interaction. The intensity of the Raman scattering as a function of the frequency of the scattered photon $\omega$ defined within an interval $d\omega$, is given~\cite{heitler1984quantum} by
\begin{equation}\label{eq:IntensityFermi}
    I(\omega) d\omega = \frac{2\pi}{\hbar} \sum_f \abs{\mel{f}{S}{i}}^2 \delta(E_f - E_i),
\end{equation}
where $\ket{i}$ and $\ket{f}$ represent the initial and final states with energies $E_i$ and $E_f$, respectively. The above formula may be extended at finite temperature but, as we will see, for our calculations the zero temperature formalism is a good descriptor. The initial state consists of a coherent state of photons with frequency $\omega_L$ (which models the incoming laser beam~\cite{siegman1986lasers}), and the summation is performed over all the final states that consist of a single photon with frequency $\omega = \omega_L - \Omega^{\textrm{tot}}_\mathrm{ph}$, since we are considering spontaneous Raman scattering, and two phonons with total frequency $\Omega^{\textrm{tot}}_\mathrm{ph}$. In fact, in spontaneous Raman scattering experiments only the frequency of the outgoing photon gets probed, while in both initial and final states the crystal electrons are in the Fermi sea. The matrix element in Eq.~\ref{eq:IntensityFermi} is given by (see Ref.~\cite{falicovmartin} and Appendix~\ref{app:matrixEl})
\begin{equation}\label{eq:FermiGoldenRule}
    \mel{f}{S}{i} = \sum_{\alpha,\beta,\gamma} \frac{\bra{f}\mathcal{H}_I\dyad{\alpha}\mathcal{H}_I\dyad{\beta}\mathcal{H}_I\dyad{\gamma}\mathcal{H}_I\ket{i}}{(\omega_\alpha - \omega_L)(\omega_\beta - \omega_L)(\omega_\gamma-\omega_L)},
\end{equation}
where $\mathcal{H}_I$ is either the $\mathcal{H}_\mathrm{el-em}$ or the $\mathcal{H}_\mathrm{el-ph}$ interaction Hamiltonian, and $\ket{\alpha}, \ket{\beta}$, and $\ket{\gamma}$ are intermediate states with energies $\omega_\alpha$, $\omega_\beta$, and $\omega_\gamma$, consisting in an arbitrary amount of photons and phonons and excited states of the electronic system. As already anticipated, the latter are commonly described in terms of electron-hole pairs, which, being only approximate eigenstates of the system, provide a non-zero imaginary part to the intermediate state energies. Moreover, the difference in the number of photons, phonons, and electron-hole pairs between the initial, intermediate, and final states is constrained by the fact that the electron-photon (electron-phonon) interaction Hamiltonian has non-zero matrix element only between states that differ by exactly one photon (phonon) and one electron-hole pair.

\begin{figure}[hbt!]
    \centering
    \includegraphics[width=\linewidth]{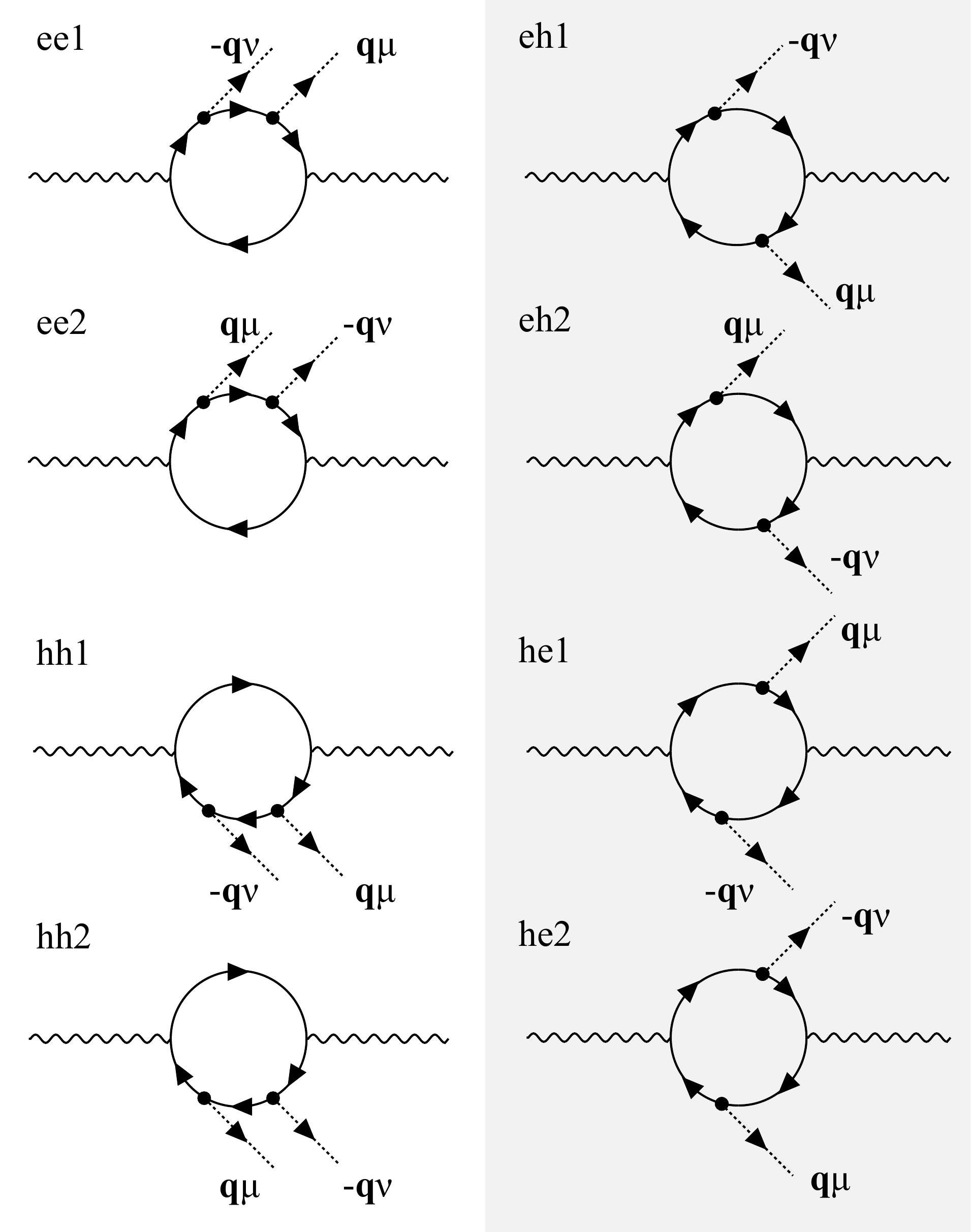}
    \caption{Feynman diagrams considered in the formalism of Ref.~\cite{venezuela2011theory} and in this work. They correspond to the $8$ resonant diagrams out of the $4! \times 3$ total ones (see App.~\ref{app:matrixEl}), where the electron-photon vertices are the first and the last to appear. The diagrams on the right (shaded region) correspond to the processes in which both the electron and the hole undergo scattering with a phonon (and they are four since we consider all the permutations of the phonon pair), and they are the dominant ones, as already shown in Ref.~\cite{venezuela2011theory} and discussed below, while the diagrams on the left (unshaded region) correspond to the processes in which both phonons arise from scattering with either the electron or the hole.}
    \label{fig:venezuela_diagrams}
\end{figure}

So far the resonance condition has not been imposed, that is no restriction has been made on the energy of the intermediate states, which may not fulfill energy conservation as a consequence of Heisenberg's uncertainty principle (so that in general they are called \emph{virtual states}). The resonance condition amounts to consider \emph{real} intermediate states, i.e.\ states which fulfill energy conservation (up to an imaginary part which describes the finite lifetime of the intermediate state), such that the real part of the factors in the denominator of Eq.~\ref{eq:FermiGoldenRule} is zero: in graphene all three intermediate states can be real states, that is the initial (final) electron-hole pair has exactly the energy of the incoming (scattered) photon, and energy is conserved also in the scattering of the electron-hole pair with the phonons. The resonance condition, which evidently can be satisfied only when the absorption (emission) of the incoming (scattered) photon is temporally ordered as the first (last) process, is responsible for the narrowness of the two-phonon Raman lines. Implementing this condition in the formalism of Eq.~\ref{eq:FermiGoldenRule} amounts to attributing the initial and final vertices of the Feynman diagrams to electron-photon interactions, and considering the $4\times2$ possible scattering processes between the two phonons and the electron-hole pair, for both $\vb{q}$ and $-\vb{q}$ phonon momentum (see Figure~\ref{fig:venezuela_diagrams}). In our calculation we will consider only these 8 diagrams.

Given the frequency $\omega_L$ of the incoming photons, the intensity of the Raman line as a function of the scattered photon frequency $\omega$ is obtained via Eq.~\ref{eq:IntensityFermi}, where as already stated one has to sum over all the possible two-phonon final states and integrate over the photon momenta. Due to crystal momentum conservation, the scattered phonons will have opposite momenta $\vb{q}$ and $-\vb{q}$, so the final states are specified by $\vb{q}$ and by the branches $\mu, \nu$ which the scattered phonons belong to. Extending Eq.~\ref{eq:IntensityFermi} to take into account finite temperature quasiparticle occupations, we obtain
\begin{equation}\label{eq:intensity_withprefactors}
    I(\omega,T) d\omega = \frac{2\pi}{\hbar} \frac{V \omega^2 d\omega}{c^3 \pi^2} N_p \tilde{I}(\omega,T),
\end{equation}
where we have explicitly integrated out the scattered photon density of states, since the photon momentum is negligible and thus does not play a role in the matrix elements, and
\begin{equation}\label{eq:RamanInt}
\begin{split}
\tilde{I}(\omega,T) = \frac{1}{N_{\vb{q}}}&\sum_{\vb{q},\mu,\nu} I_{\nu\mu}(\vb{q}) \delta\left(\omega_L - \omega - \omega_{-\vb{q}}^\nu - \omega_{\vb{q}}^\mu\right) \times \\
&\times \left[n(\omega_{\vb{-q}}^\nu)+1\right]\left[n(\omega_{\vb{q}}^\mu)+1\right]
\end{split}
\end{equation}
where $n(\omega_{\vb{q}}^\mu)$ is the Bose-Einstein distribution function, i.e.\ the occupation number of the phonon state having energy $\omega_{\vb{q}}^\mu$. At room temperature, for the phonons in which we are interested in, $\hbar\omega_{\vb{q}}^\mu\gg k_B T$, thus $n(\omega_{\vb{q}}^\mu) \simeq 0$ and $I(\omega,T)\sim I(\omega)$ as anticipated. The total energy conservation between the initial and final state is enforced by the Dirac delta function, but since the final pair of phonons are quasiparticles and thus possess a finite lifetime, we substitute it with a Lorentzian function having the width equal to the inverse lifetime of the final phonon pair (which is essentially due to anharmonicity, see below). The square modulus of the matrix element $\mel{f}{S}{i}$ of Eq.~\ref{eq:IntensityFermi} becomes $I_{\nu\mu}(\vb{q})$, defined as
\begin{equation}\label{eq:Iq}
I_{\nu\mu}(\vb{q}) = \abs\Big{\frac{1}{N_{\vb{k}}} \sum_{\vb{k},\beta} K_\beta(\vb{k},\vb{q},\nu,\mu)}^2,
\end{equation}
where the expression for $K_{\beta}(\mathbf{k},\mathbf{q},\nu,\mu)$ can be deduced from Eq.~\ref{eq:FermiGoldenRule} and is detailed in Ref.~\cite{venezuela2011theory}.
Since all the resonant processes happens between the $\pi$ and $\pi^*$ electronic bands, and given the constraint on the difference of the number of photons and phonons between intermediate states which we discussed above, to specify the intermediate state it is sufficient to indicate the momentum $\vb{k}$ of the electron-hole pair and the momentum $\vb{q}$ and branch indexes $\nu$ and $\mu$ of the phonons. $\beta$ labels instead the $8$ different time-orderings of the electron/hole-phonon scattering processes represented by the diagrams of Fig.~\ref{fig:venezuela_diagrams}.

For the sake of completeness, we specify that in the limit of infinite volume $V$ Eq.~\ref{eq:intensity_withprefactors} gives formally zero, since, as it will be detailed below, the matrix element $\mel{f}{S}{i}$ which appears squared inside $\tilde{I}(\omega)$ scales as $1/V$. One has indeed to consider the scattering cross-section, which is defined~\cite{heitler1984quantum} as 
\begin{equation}\label{eq:crossSec}
    d\sigma = \frac{V}{\bar{n} c} I(\omega) d\omega,
\end{equation}
where $\bar{n}$ is the average number of photons in the laser's coherent state, and which has the dimensions of a squared length, $I(\omega)$ being adimensional. This corresponds experimentally to the proportionality factor between the intensity of the scattered photons and the power flux of incident laser photons.

\subsubsection{Matrix elements}\label{subsubsec:mel}
We proceed with the evaluation of the matrix elements $\mel{\alpha}{\mathcal{H}_I}{\beta}$ of Eq.~\ref{eq:FermiGoldenRule}, where $\mathcal{H}_I$ indicates either $\mathcal{H}_\mathrm{el-ph}$ or $\mathcal{H}_\mathrm{el-em}$. The initial state of the system is given by $\ket{\mathrm{eGS}}(\ket{\bar{n}}^\mathrm{coh}_{\omega_L}\ket{0}_{\omega})\ket{0}$, where the electromagnetic field is described by the product of a coherent state with $\bar{n}$ average photons at the laser frequency $\omega_L$ with wavevector $\vb{p}_L$, and a Fock state of zero photons at the frequency $\omega$, while the final state is given by $\bra{\mathrm{eGS}}(\bra{\bar{n}}^\mathrm{coh}_{\omega_L}\bra{1}_{\omega})\bra{\omega_{\vb{q}}^\nu, \omega_{-\vb{q}}^\mu}$, where there is instead a photon of frequency $\omega$ in addition to the laser's coherent state and two phonons of frequencies $\omega_{\vb{q}}^\nu$ and $\omega_{-\vb{q}}^\mu$, having opposite wavevectors $\vb{q}$ and belonging to branches $\nu$ and $\mu$, respectively.

The matrix element for the interaction of the incoming laser beam with the electronic degrees of freedom is given by  
\begin{equation}
\begin{split}
        \mel{\gamma}{\mathcal{H}_\mathrm{el-em}}{i} = - \frac{e}{m} \sqrt{\frac{\hbar\bar{n}}{2\varepsilon_0 V \omega_L}} \times \\ 
        \times \sum_{\substack{\vb{k}, \alpha \\ \vb{k'}, \alpha'}} \hat{\epsilon}_{\vb{p}_L}^{\text{in}}
        \cdot \mel{\vb{k}, \alpha}{\vb{p}\, e^{i \vb{p}_L\cdot \vb{r}}}{\vb{k'}, \alpha'} \mel{\gamma}{c_{\vb{k}}^{\alpha\dagger} c_{\vb{k'}}^{\alpha'}}{\mathrm{eGS}}
        \label{eq:p1}
\end{split}
\end{equation}
where the polarization vector is assumed to lie in the graphene $(x,y)$ plane, remembering  that in the Coulomb gauge $\hat{\epsilon}_{\vb{p}_L} \cdot \vb{p}_L = 0$. Notice that the factor $\sqrt{\bar{n}}$, once the matrix element is squared, will simplify with the photon flux of Eq.~\ref{eq:crossSec}, so that effectively one could describe the interaction with one incoming photon only. Being the coherent state an eigenstate of the annihilation operator, the electromagnetic field of the $\ket{\gamma}$ state is described by the same laser's coherent state, and in addition $\ket{\gamma}$ contains an electron-hole pair which, due to the fact that the photon wavevector is negligible, has zero total momentum ($\vb{k} = \vb{k'}$); we also restrict to the resonant case $\alpha'=\pi$ and $\alpha=\pi^*$. From now on, we will drop the momentum index for the photon states. In complete analogy one can obtain the matrix element for the interaction of the scattered photon with the electronic degrees of freedom
\begin{equation}
\begin{split}
    \mel{f}{\mathcal{H}_\mathrm{el-em}}{\gamma} = - \frac{e}{m} \sqrt{\frac{\hbar}{2\varepsilon_0 V \omega}} \times \\ \times \sum_{\vb{k}} \hat{\epsilon}^{\text{out}} \cdot \mel{\vb{k}, \pi}{\vb{p}}{\vb{k}, \pi^*}.
    \label{eq:p2}
\end{split}
\end{equation}
Notice that when a non-local pseudo-potential is used to approximate the electron-ion interaction in the Hamiltonian, such as in the \emph{ab initio} calculations performed in the present work, the matrix element of the momentum operator of Eqs.~\ref{eq:p1} and~\ref{eq:p2} must be replaced by the matrix element of the commutator between the Hamiltonian and the position operator~\cite{starace1971length,PhysRevB.44.13071}.
The matrix element for the electron-phonon interaction is readily given by Eq.~\ref{eq:EPC}, that is
\begin{equation}
    \mel{\beta}{\mathcal{H}_\mathrm{el-ph}}{\gamma} = g_{\vb{k+q}\pi/\pi^*,\vb{k}\pi/\pi^*}^\nu,
\end{equation}
which describes the scattering of an electron (or a hole) from the state with wavevector $\vb{k}$ in band $\pi/\pi^*$ to the state with wavevector $\vb{k+q}$ with the same band index \cite{venezuela2011theory}, with the simultaneous emission of a phonon with wavevector $\vb{q}$ belonging to branch $\nu$. Again in either the intermediate state $\ket{\beta}$ or $\ket{\gamma}$ the electromagnetic field is in the state $\ket{\bar{n}}^\mathrm{coh}_{\omega_L}\ket{0}_{\omega}$. In the following we will deal with phonons belonging to the TO branch, which is the highest-optical branch at \textbf{K}~\cite{gruneis2009phonon}, with wavevector $\vb{q}$ in the vicinity either of the \textbf{K} or the $\bm{\Gamma}$ point, so we will omit the branch index $\nu$ when possible. It will also prove useful in the following to define, as in Ref.~\cite{PhysRevLett.95.236802}, $\expval{D^2_{\bm{\Gamma}}}$ and $\expval{D^2_{\mathrm{\textbf{K}}}}$ as the average square of $D_{\vb{k+q}\pi/\pi^*,\vb{k}\pi/\pi^*}$ between electronic states at the resonance wavevectors $\vb{k} \sim \mathrm{\textbf{K}}$, for $\vb{q} = \bm{\Gamma}$ and $\vb{q} = \mathrm{\textbf{K}}$, respectively. We further define $\expval{g^2_{\bm{\Gamma}}} = \big(\hbar/(2M\omega_{\bm{\Gamma}}) \big) \expval{D^2_{\bm{\Gamma}}}$ and $\expval{g^2_\mathrm{\textbf{K}}} = \big( \hbar/(2M\omega_\mathrm{\textbf{K}}) \big) \expval{D^2_{\mathrm{\textbf{K}}}}$. 

Since the ratio between the EPC at \textbf{K} and at $\bm{\Gamma}$ is experimentally related to the ratio of the integrated areas under the 2D and 2D' resonance peaks, we further define
\begin{equation}\label{eq:rvc}
    r_\mathrm{vc} = \frac{\expval{D^2_{\mathrm{\textbf{K}}}}}{2\expval{D^2_{\bm{\Gamma}}}} = \frac{\expval{g^2_\mathrm{\textbf{K}}}}{2 \expval{g^2_{\bm{\Gamma}}} }\frac{\omega_\mathrm{\textbf{K}}}{\omega_{\bm{\Gamma}}},
\end{equation}
which evaluates to about one if all the ingredients are computed in DFT~\cite{piscanec2004kohn}. The subscript $_\mathrm{vc}$ stands for ``vertex correction'', since in Ref.~\cite{graziotto2022probing} the failure of DFT in explaining the strong enhancement of the EPC is attributed to the neglect of the Coulomb vertex corrections, and $r_\mathrm{vc}$ will be employed as a rescaling factor that keeps them into account.

\subsubsection{Resonance conditions in the conical model without matrix elements}
\begin{figure*}
    \centering
    \includegraphics[width=0.6\linewidth]{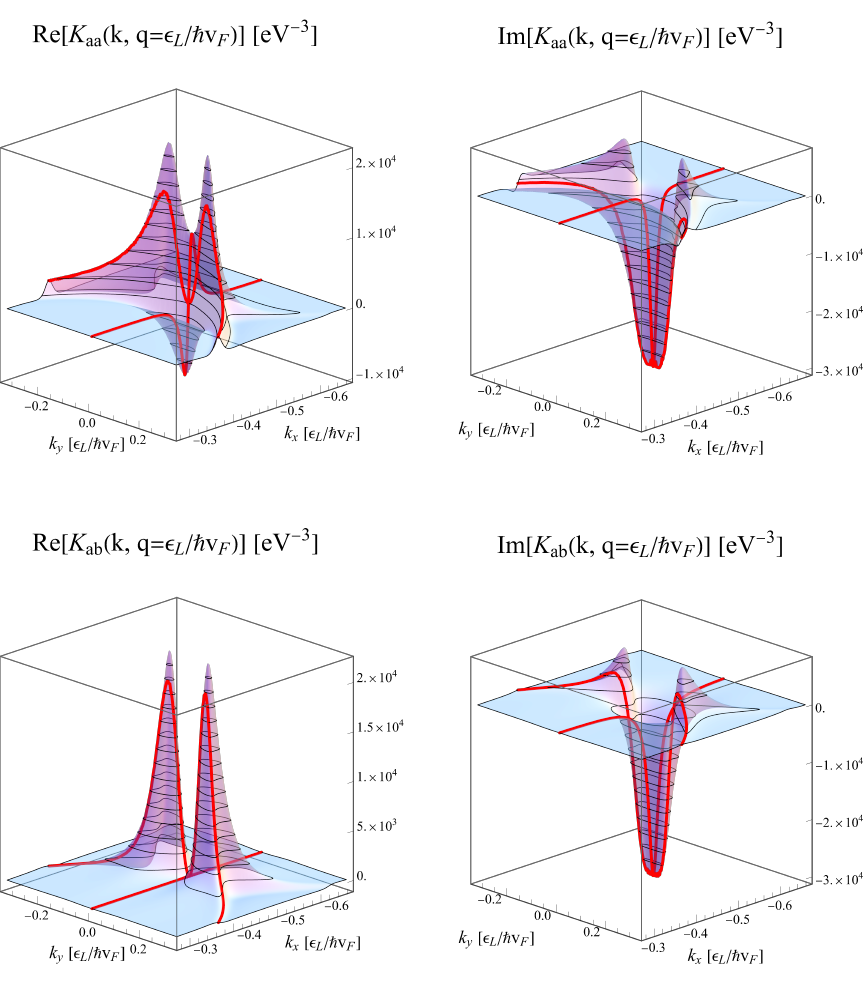}
    \caption{Real and imaginary part of the amplitudes $K_\mathrm{aa}(\vb{k},\vb{q})$, $K_\mathrm{ab}(\vb{k},\vb{q})$ defined in Eqs.~\ref{eq:Kaa}, \ref{eq:Kab} evaluated at $\vb{q} = (\epsilon_L / \hbar v_F , 0)$ for the case $\omega_\mathrm{\textbf{K}} = v_\mathrm{ph} = 0$, with $\epsilon_L = \SI{1.16}{eV}$ and $\gamma = \SI{64}{meV}$, the latter being double the value used in the calculations below for illustrative purposes. The simultaneous vanishing of the real parts of the factors at the denominator (i.e.\ the resonance condition) happens at the intersection of the two red lines, which mark the direction $\vb{k} = -\vb{q} / 2$  and the circumference $\abs{\vb{k}} = \abs{\vb{q}}/2$.  Notice how the real part of $K_\mathrm{aa}$ changes sign near the resonant value $\vb{k} = -\vb{q} / 2$, while the real part of $K_\mathrm{ab}$ does not. Moreover, notice the asymmetric shape of the peak in the imaginary part of $K_\mathrm{aa}$, as opposed to the symmetric shape of the imaginary part of $K_\mathrm{ab}$. We point out that in the limit $\gamma \to 0$ the two peaks which one can distinguish in the panels superimpose, and their values diverge.}
    \label{fig:Kaa_Kab_Re_Im} 
\end{figure*}

To qualitatively explain how the resonance condition arises in the scattering amplitudes $K_\beta$ of Eq.~\ref{eq:Iq} we consider a simplified conical description of the graphene bands,
\begin{equation}
    \epsilon_{\vb{k}}^{\pi^* / \pi} = \pm \hbar v_F \abs{\vb{k}},
\end{equation}
where $v_F$ is the Fermi velocity evaluated via GW calculations~\cite{hedin1965new} ($\hbar v_F = \SI{6.44}{eV \text{\AA}}$) and $\vb{k}$ is measured from the \textbf{K}. For the 2D peak (the case of the 2D' peak is analogous) we have for the TO phonon dispersion,
\begin{equation}
    \omega_{\vb{q}} = \omega_\mathrm{\textbf{K}} + \hbar v_\mathrm{ph} \abs{\vb{q}},
\end{equation}
where $\omega_\mathrm{\textbf{K}}$ is the phonon frequency at the \textbf{K} point given in Table~\ref{tab:gamma_params}, $v_\mathrm{ph}$ is the slope of the Kohn anomaly obtained in Ref.~\cite{piscanec2004kohn} ($\hbar v_\mathrm{ph} = \SI{0.047}{eV \text{\AA}}$), and again $\vb{q}$ is measured from \textbf{K}. Moreover, we consider neither the electron-phonon nor the electron-light matrix elements, that is we set the numerator of Eq.~\ref{eq:FermiGoldenRule} equal to one. In this approximation, the number of independent diagrams of Fig.~\ref{fig:venezuela_diagrams} reduces to 2, and their amplitude is given by the following expressions
\begin{align}
\begin{split}\label{eq:Kaa}
            & K_\mathrm{aa}(\vb{k},\vb{q}) = \frac{1}{\epsilon_L - 2\hbar v_F \abs{\vb{k}} - 2\omega_\mathrm{\textbf{K}} - 2\hbar v_\mathrm{ph} \abs{\vb{q}} - i\gamma/2} \\ &\times \frac{1}{\epsilon_L - \hbar v_F \abs{\vb{k+q}} - \hbar v_F \abs{\vb{k}} - \omega_\mathrm{\textbf{K}} - \hbar v_\mathrm{ph} \abs{\vb{q}} - i \gamma / 2} \\ &\times \frac{1}{\epsilon_L - 2\hbar v_F \abs{\vb{k}} - i\gamma/2}\\
\end{split}
\end{align}
\begin{align}
\begin{split}\label{eq:Kab}
           & K_\mathrm{ab}(\vb{k},\vb{q}) = \frac{1}{\epsilon_L - 2\hbar v_F \abs{\vb{k+q}} - 2\omega_\mathrm{\textbf{K}} - 2\hbar v_\mathrm{ph} \abs{\vb{q}} - i\gamma/2} \\ &\times \frac{1}{\epsilon_L - \hbar v_F \abs{\vb{k+q}} - \hbar v_F \abs{\vb{k}} - \omega_\mathrm{\textbf{K}} - \hbar v_\mathrm{ph} \abs{\vb{q}} - i \gamma / 2} \\ &\times \frac{1}{\epsilon_L - 2\hbar v_F \abs{\vb{k}} - i\gamma/2}, 
\end{split}
\end{align}
where $\gamma$ is the total inverse lifetime of the intermediate states (which is given by twice the inverse lifetime $\gamma^{e/h}$ of the electron/hole, supposed equal, where $\gamma^{e/h}$ is the full-width at half maximum (FWHM) of the electronic spectral function, or equivalently minus two times the imaginary part of the electronic self-energy), and the subscripts `aa' and `ab' refer to the four diagrams on the left or on the right of Fig.~\ref{fig:venezuela_diagrams}, respectively. The intensity as a function of phonon wavevector $\vb{q}$ is given by Eq.~\ref{eq:Iq} to be
\begin{equation}\label{eq:Iq_conical}
    I(\vb{q}) = \abs\Big{\frac{1}{N_{\vb{k}}} \sum_{\vb{k}} \big( K_\mathrm{aa}(\vb{k},\vb{q}) + K_\mathrm{ab}(\vb{k},\vb{q})\big)}^2.
\end{equation}
Let us first consider the case in which $\omega_\mathrm{\textbf{K}} = v_\mathrm{ph} = 0$: the resonance condition corresponds to the vanishing of the real part of the three factors in the denominators of Eq.~\ref{eq:Kaa}, \ref{eq:Kab}, which can simultaneously happen for $\vb{q} = -2\vb{k}$, $\abs{\vb{k}} = \epsilon_L / (2\hbar v_F)$, giving rise to a triple resonance condition. However, the behaviour of $K_\mathrm{aa}$ and $K_\mathrm{ab}$ near the resonance is quite different (see Fig.~\ref{fig:Kaa_Kab_Re_Im}), with the real part of $K_\mathrm{aa}$ changing sign along the resonance region, while the real part of $K_\mathrm{ab}$ stays positive. This implies, as already shown in Ref.~\cite{venezuela2011theory}, that the predominant contribution to the Raman intensity is given by $K_\mathrm{ab}$, which add up coherently in a constructive way when summing over the electronic wavevectors $\vb{k}$ in Eq.~\ref{eq:Iq_conical}, at variance with the destructive interference of the terms $K_\mathrm{aa}$. When a finite value is considered for both $\omega_\mathrm{\textbf{K}}$ and $v_\mathrm{ph}$ this difference is made even stronger by the fact that in Eq.~\ref{eq:Kaa} one cannot achieve anymore the fully triple resonance condition~\cite{basko2008theory} (i.e.\ there is a double resonance at most).

Limiting then ourselves to the `ab' process only, we obtain an analytical expression for $I(\vb{q})$, which is formally identical to the square modulus of Eq.~63 of Ref.~\cite{basko2008theory}, as it indeed should be since we are eventually considering the same processes, even if the two formalisms are different (one has to remember the different definition of the total inverse lifetime $\gamma$, which here is four times the $\gamma$ of Ref.~\cite{basko2008theory}, since there it refers directly to the negative imaginary part of the electronic self-energy):
\begin{equation}
    I(\vb{q}) = \frac{\omega_L}{2^7 v_F^4} \left[\Big(\omega_L - \omega_\mathrm{\textbf{K}} - (v_F + v_\mathrm{ph}) q\Big)^2 + \left(\frac{\gamma}{2}\right)^2 \right]^{-3/2},
\end{equation}
which is peaked at $q \equiv \abs{\vb{q}} = (\omega_L - \omega_\mathrm{\textbf{K}}) / (v_F + v_\mathrm{ph})$, and it has full width at half maximum
\begin{equation}
    \mathrm{FWHM} = \frac{\sqrt{2^{2/3}-1}}{\hbar (v_F + v_\mathrm{ph})} \gamma.
\end{equation}
The analytical result is obtained by approximating the summation on $\vb{k}$ in Eq.~\ref{eq:Iq_conical} with an integral about the resonance contour $\abs{\vb{k}} \simeq \omega_L / (2 v_F)$. Notice that including the phononic dispersion amounts simply to substitute $v_F \rightarrow v_F + v_\mathrm{ph}$ in the denominator of Eq.~63 of Ref.~\cite{basko2008theory}.
Neglecting for simplicity the finite lifetime of the final phonon states, we may further obtain the intensity as a function of the scattered photon energy as
\begin{equation}\label{eq:Baskovian}
    I(\omega) \propto \left[\Big(\frac{v_F}{v_\mathrm{ph}}\Big)^2 \Big(\frac{\epsilon_L - \hbar \omega}{2} - \omega_\mathrm{ph} \Big)^2 + \Big(\frac{\gamma}{2}\Big)^2  \right]^{-3/2},
\end{equation}
where we recognize the \emph{Raman shift} as $\epsilon_L - \hbar\omega$, which is peaked at $\omega=2\omega_{\textrm{ph}}$ where $\omega_\mathrm{ph} = \omega_\mathrm{\textbf{K}} + \hbar v_\mathrm{ph}(\omega_L - \omega_\mathrm{\textbf{K}}) / (v_F + v_\mathrm{ph})$.

The function described by Eq.~\ref{eq:Baskovian} has the form
\begin{equation}\label{eq:Baskovianform}
    f_B(\omega) = I \frac{\beta^3}{\big( (\omega-\omega_0)^2 + \beta^2\big)^{3/2}},
\end{equation}
where $I$ is a coefficient which contains the fourth power of the EPC at \textbf{K} ($\boldsymbol{\Gamma}$), $\omega_0$ is the central frequency of the peak, equal to twice the value of the phonon energy at the resonance wavevector near \textbf{K} ($\boldsymbol{\Gamma}$), and the FWHM is given by 
\begin{equation}\label{eq:FWHM_Baskovian}
    \mathrm{FWHM}_B = 2\sqrt{2^{2/3}-1}\,\beta \equiv 2\sqrt{2^{2/3}-1}\,\gamma
    \frac{v_\mathrm{ph}}{v_F}.
\end{equation}
The functional form $f_B(\omega)$ will be referred to as \emph{Baskovian} in the following, since it was first calculated by D.\ Basko in Eq.~2 in Ref.~\cite{basko2008theory}. One may further calculate the integrated area under $f_B(\omega)$, which gives
\begin{equation}\label{eq:A_Baskovian}
    \mathrm{A}_B = 2 I \beta.
\end{equation}
Taking into account all the proportionality factors, the coefficients which appear in the matrix elements, and the factors due to the final photon density of states (see Eq.~\ref{eq:intensity_withprefactors}), the integrated area under the 2D peak in the conical model approximation is given by Eq.~66 of Ref.~\cite{basko2008theory}: here we notice in particular that it scales as $(\omega_L - 2\omega_\mathrm{ph})^2 / \gamma^2$. Since, as it will be shown below, $\gamma$ scales linearly with $\omega_L$, it follows that the integrated area is almost constant as a function of $\epsilon_L$. 

\section{Computational Approach}\label{sec:Comp_Appr}
The computational infrastructure used to compute the diagrams of Fig.~\ref{fig:venezuela_diagrams} is the \texttt{EPIq} code~\cite{marini2023epiq}. We now detail the computational parameters used to determine all the physical quantities presented in Sec.~\ref{sec:theo_frame}.
\subsection{Electronic states}
\label{sec:elecstates}
\begin{figure}[hbt!]
    \centering
    \includegraphics[width=0.9\linewidth]{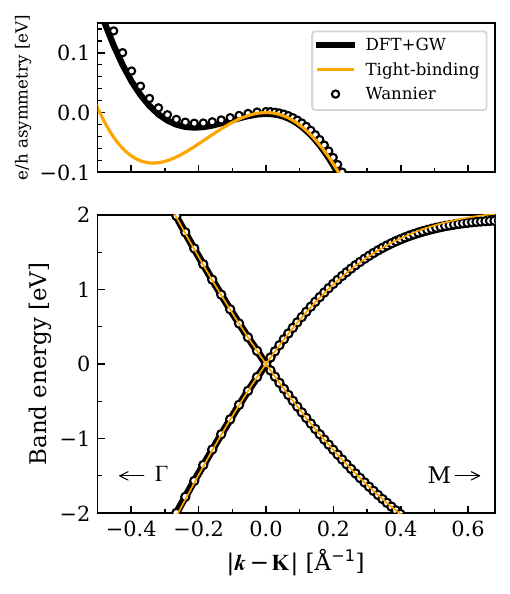}
    \caption{Comparison between the electronic dispersion calculated with five-neighbours tight-binding in Ref.~\cite{venezuela2011theory} (orange curve), the dispersion calculated within DFT multiplied by the GW correction factor (black curve), and the dispersion which we employed in this work, resulting from the wannierization procedure (open circles), along the high symmetry $\bm{\Gamma}$-\textbf{K}-\textbf{M} line.}
    \label{fig:TB_DFT_electrons}
\end{figure}

The Kohn-Sham states and eigenenergies are obtained within DFT using Quantum Espresso (QE)~\cite{giannozzi2009quantum}, by modelling the monolayer graphene honeycomb structure with two carbon atoms per unit cell (with four valence electrons each) and lattice parameter $a=\SI{2.46}{\text{\AA}}$, on a $64\times64$ electronic grid. From the energetically lowest ten electronic bands, we extract maximally localized Wannier functions~\cite{marzari2003introduction} (MLWF) using the Wannier90 (W90) software~\cite{mostofi2008wannier90}. As anticipated, the $\pi^*$ and $\pi$ bands  are the only ones considered in the Raman intensity calculation. We multiply them by a corrective factor 1.18 (after setting the Fermi energy to zero) to reproduce the band energy slope obtained from GW calculations, and which shows the best agreement with angular resolved photo-emission spectroscopy (ARPES) measurements~\cite{gruneis2008tight}. As compared to the five nearest-neighbour tight-binding approach employed in Ref.~\cite{venezuela2011theory}, the approach we employed in this work is better at reproducing the trigonal warping and the electron-hole asymmetry of the electronic dispersion. Most importantly, it enables the use of MLWFs, which are better suited to faithfully reproduce the matrix elements dependence in $\mathbf{k}$ space rather than simplified models. In Figure~\ref{fig:TB_DFT_electrons} we show a comparison of our GW-corrected DFT result and of the tight-binding dispersion of Ref.~\cite{venezuela2011theory}, along the high-symmetry $\bm{\Gamma}$-\textbf{K}-\textbf{M} line. The fine momentum grids used for the Wannier interpolation of electronic properties are built using a ``telescopic'' procedure, and are different for every excitation energy, as described in Appendix~\ref{app:telescopic}.

\subsection{Phononic states}\label{subsec:phonons}

\begin{figure}[hbt]
    \centering
    \includegraphics[width=0.9\linewidth]{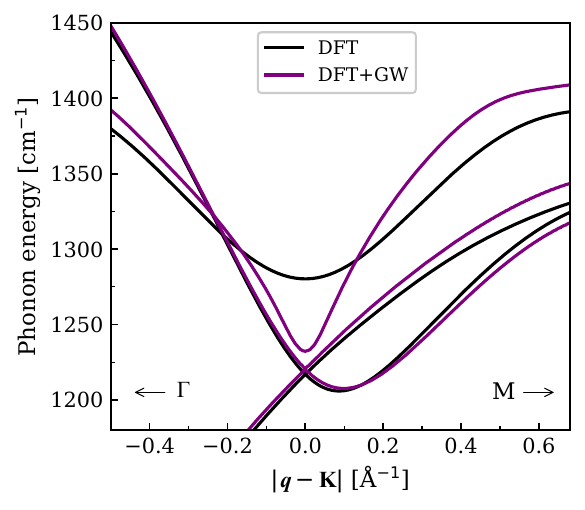}
    \caption{Phonon dispersion along the high-symmetry $\bm{\Gamma}$-\textbf{K}-\textbf{M} line. The three highest branches are shown and they are the result of DFT calculation (black curves), and DFT corrected with GW with the methodology of Ref.~\cite{graziotto2022probing, herziger2014two} (purple curves). Notice how DFT alone is not able to reproduce correctly the Kohn anomaly at \textbf{K}.}
    \label{fig:DFT_GW_phonons}
\end{figure}

The phonon dispersion of graphene computed in DFT wrongly underestimates the slope of the Kohn anomalies at the \textbf{K} and \textbf{K'} points (associated to the transverse optical (TO) mode \cite{gruneis2009phonon}, see Fig.~\ref{fig:DFT_GW_phonons}) by a factor of 2~\cite{lazzeri2008impact}. Indeed, as it was shown in Ref.~\cite{piscanec2004kohn}, the Kohn anomaly is entirely determined by the contribution of the phonon self-energy between the electronic $\pi$ and $\pi^*$ bands. Therefore, following the same procedure of Ref.~\cite{herziger2014two}, the dynamical matrix $\mathcal{D}_{\vb{q}}$ is first calculated with linear response in DFT on a $6\times6$ uniform grid in the FBZ, then it is Fourier interpolated on a uniform finer $400\times400$ grid in the FBZ, and finally the GW correction is applied to the TO mode. The details of the phonon dispersion, such as the slope of the Kohn anomaly and the trigonal warping, bear a strong influence on the Raman spectrum: for instance, as it will be discussed below, the trigonal warping of the phonon dispersion is crucial in determining the shape of the 2D peak, while it does not influence the total integrated area under the peak. We remark in particular that the phonon dispersion plays a negligible role in the denominators of the amplitudes $K_\beta(\vb{k},\vb{q},\nu,\mu)$, i.e.\ we could replace it with a constant energy equal to the TO frequency at the edge of the FBZ with negligible error; instead, it is crucial to consider it in the delta function of Eq.~\ref{eq:RamanInt}. As it was mentioned above, the finite lifetime of the phonon states is taken into account only by replacing the delta with a Lorentzian function with full-width at half maximum (FWHM) \SI{5.3}{cm^{-1}}, given by twice the value of the linewidth of the TO phonons \cite{paulatto2013anharmonic} near the \textbf{K} point, since TO phonons are responsible for the 2D and 2D' peaks. 

\subsection{Electron-light scattering}
The electron-light interaction is obtained via Wannier interpolation of the unscreened electric dipole calculated within the LDA approximation, following the same procedure of Ref.~\cite{herziger2014two}. We assume that the polarization of the incoming and scattered light lies on the graphene $(x,y)$ plane, and although one can resolve both the incoming and outgoing light polarization (as it has been done in Ref.~\cite{popov2012theoretical}) we will display results obtained considering unpolarized laser excitation and summing over all the possible polarizations of the scattered light
\begin{equation}
I_\mathrm{unpol} = \frac{1}{2} \sum_i \abs{\sum_o A_{i,o}}^2,
\end{equation}
where $i,o = x,y$ label the polarization of the incoming and outgoing light, respectively, while $A_{i,o}$ represents the scattering amplitude, i.e.\ the argument of the absolute value in Eq.~\ref{eq:Iq}. This formula can be simplified in the following way: suppose that the impinging light has polarization $\hat{\mathbf{e}}=\cos(\theta) e_x+ \sin(\theta) e_y$, while the outgoing light has polarization $\hat{\mathbf{e}}=\cos(\phi) e_x+ \sin(\phi) e_y$. When performing the square modulus of the sum of the amplitudes, we will have terms proportional to $\cos^2(\theta)\cos^2(\phi),\cos^2(\theta)\cos(\phi)\sin(\phi),\ldots$ ; if we now assume that we are resolving the intensity over a time period at the condition that the impinging light is unpolarized, this is equivalent to average $\theta$ over a uniform distribution, and integrating over $\phi$. In such case, the only terms which contribute to the intensity are the ones containing even powers of the trigonometric functions, i.e.\ $I_{xx},I_{xy},I_{yx},I_{yy}$. Hence
\begin{equation}
    I_\mathrm{unpol} = \frac{1}{2} \big( I_{xx} + I_{xy} + I_{yx} + I_{yy} \big).
\end{equation}

\subsection{Electron-phonon scattering}
The EPC matrix element in the Bloch basis set is defined as in Eq.~\ref{eq:EPC}, and computed \emph{ab initio}/interpolated on the same fine grids described in Secs.~\ref{sec:elecstates} and \ref{subsec:phonons}. The GW correction to the TO mode over the whole FBZ are included following Ref.~\cite{venezuela2011theory}, i.e.\ considering the modification of the phononic frequency and polarization vector. 


\subsection{Electron-hole linewidth}\label{subsubsec:elec_hole_gamma}
As already mentioned above, the description of the excited states of monolayer graphene in terms of creation and annihilation of electron-hole pairs is an approximated one, and does not provide access to the real spectrum of the system. Indeed, the quasi-particle electronic states $\ket{\vb{k},\alpha}$ do possess a finite lifetime $\tau_{\vb{k}}^\alpha$ (or analogously a non-zero FWHM linewidth $\gamma_{\vb{k}}^\alpha = \hbar/\tau_{\vb{k}}^\alpha$) because they interact with phonons, with other electronic states, or with defects. One can directly measure the linewidth $\gamma_{\vb{k}}^\alpha$ as the FWHM of the electron/hole spectral function, e.g.\ with ARPES, which has a crucial role in determining the line-shape of the double-resonance Raman peaks. In fact, neglecting inhomogeneous broadening which can arise from fluctuations of the strain of the sample~\cite{neumann2015raman}, there are two sources of homogeneous broadening of the line-shape: phononic, which as discussed in Section~\ref{subsec:phonons} we consider only through the delta function of Eq.~\ref{eq:RamanInt}, and electronic, which depends on $\gamma_{\vb{k}}^\alpha$. We neglect the contribution due to electron-defect scattering, which is suppressed in pristine monolayer samples, and the electron-electron scattering contribution, which for undoped graphene is proven to be negligible~\cite{basko2009electron}. According to Fermi's golden rule, the electron-phonon scattering contribution is given by Ref.~\cite{venezuela2011theory} to be:
\begin{equation}
\begin{split}
    \gamma_{\vb{k}}^{\alpha(\mathrm{e-ph})} = & 2\pi \int \frac{\dd^2q}{(2\pi)^2}\sum_\nu \abs{\mel{\vb{k}+\vb{q},\alpha}{\Delta\mathcal{H}_{\vb{q},\nu}}{\vb{k},\alpha}}^2 \\
    &\times \delta(\epsilon_{\vb{k}}^\alpha - \epsilon_{\vb{k}+\vb{q}}^\alpha - \hbar\omega_{-\vb{q}}^\nu),
\end{split}
\end{equation}
where the integration is performed over the FBZ and the summation over all phonon branches $\nu$. Notice that the electron-phonon scattering does not change the electronic band $\alpha=\pi,\pi^*$. Considering conical bands and only the two phonons at $\bm{\Gamma}$ and at \textbf{K} one obtains (see Ref.~\cite{venezuela2011theory}):
\begin{equation}\label{eq:elec_hole_gamma}
\begin{split}
      \gamma_{\mathrm{conical}}^{\alpha(\mathrm{e-ph})}(\epsilon_L) = \frac{\pi}{2} \bigg[& 2 \expval{g^2_{\bm{\Gamma}}} N_a\left(\frac{\epsilon_L}{2} - \hbar \omega_{\bm{\Gamma}}\right)  + \\
    &+ \expval{g^2_{\mathrm{\textbf{K}}}} N_a\left(\frac{\epsilon_L}{2} - \hbar \omega_\mathrm{\textbf{K}}\right)  \bigg],  
\end{split}
\end{equation}
\begin{equation}
    N_a(\epsilon) = \frac{\sqrt{3}}{\pi} \left(\frac{a_0}{\hbar v_F}\right)^2 \abs{\epsilon} \theta(\abs{\epsilon}),
\end{equation}
where $a_0 = \SI{2.46}{\text{\AA}}$ is the lattice constant and $\hbar v_F = \SI{6.44}{eV \text{\AA}}$. 
Using the values given in Table~\ref{tab:gamma_params}, the inverse lifetime of the electron/hole is given by the same result of Ref.~\cite{venezuela2011theory} to be
\begin{equation}\label{eq:gamma_Pedro}
    \gamma_{\mathrm{conical}}^{\alpha(\mathrm{e-ph})}(\epsilon_L) = 41.35(\epsilon_L/2-0.166) \, \SI{}{meV},
\end{equation}
where $\epsilon_L$ is in \SI{}{eV}. The inverse lifetime that appears in the denominator of $K_\beta(\vb{k},\vb{q},\nu,\mu)$ is the inverse lifetime of the total state which, neglecting the phonon contribution, is the sum of the electron and hole lifetimes. Supposing electron-hole symmetry and neglecting the dependence on $\vb{k}$, as in Ref.~\cite{venezuela2011theory} the total electron-hole FWHM linewidth reads
\begin{equation}\label{eq:gamma_tot}
    \gamma_{\mathrm{tot}} \equiv \gamma_{\vb{k}}^\alpha = \gamma_{\vb{k}}^\beta = \gamma_{\vb{k}}^\gamma = 2\times\gamma_{\mathrm{conical}}^{\alpha(\mathrm{e-ph})},
\end{equation}
where the superscripts $\alpha, \beta, \gamma$ label the intermediate states (see Eq.~\ref{eq:FermiGoldenRule}).
Being dependent on the EPC, the electronic linewidth is also affected by the rescaling factor $r_\mathrm{vc}$, as discussed in the following.

\begin{table}[!hbt]
    \centering
    \begin{tabular}{c|c|c|c}
         & $\expval{D^2_{\vb{q}}}$ (\SI{}{eV^2 / \text{\AA}^2}) & $\omega_{\vb{q}}$ (\SI{}{cm^{-1}}) & $\expval{g^2_{\vb{q}}}$ (\SI{}{eV^2}) \rule[-1.2ex]{0pt}{0pt} \\
          \hline
        $\bm{\Gamma}$ & $62.8$ & $1580$ & $0.056$ \\
        \textbf{K} & $193$ & $1220$ & $0.222$
    \end{tabular}
    \caption{Values employed in the calculation of the inverse lifetime of the electronic states. $\expval{D^2_{\bm{\Gamma}/\mathbf{K}}}$ are given in Ref.~\cite{lazzeri2008impact}, and by using $\omega_{\bm{\Gamma},\mathbf{K}}$ and the definition of Sec.~\ref{subsubsec:mel} we obtain $\expval{g^2_{\bm{\Gamma}/\mathbf{K}}}$. The values given in this table will be used also for later calculations, if not specified differently.}
    \label{tab:gamma_params}
\end{table}

\section{Results and discussion}\label{sec:Res_Disc}
\begin{figure}[hbt!]
    \centering
    \includegraphics[width=0.9\linewidth]{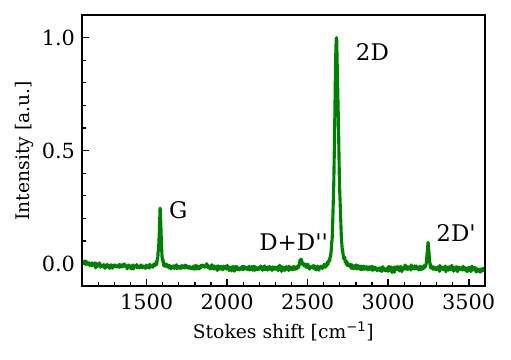}
    \caption{Typical Raman spectrum of monolayer graphene, measured with \SI{2.33}{eV} excitation energy. The letters refer to the common nomenclature of the Raman peaks. Notice the non-resonant first-order G peak at a frequency of \SI{1584}{cm^{-1}}, which is due to the iTO and iLO phonons at $\bm{\Gamma}$, and the resonant second-order narrow peaks (D+D'', 2D and 2D'). The spectrum has been normalized to the intensity of the 2D peak.}
    \label{fig:ramanSpectrum_Exp}
\end{figure}

The experimental Raman spectrum of graphene is displayed in Figure~\ref{fig:ramanSpectrum_Exp}: one can recognize the non-resonant first order G peak at \SI{1584}{cm^{-1}} Stokes shift, which as already explained above is due to the degenerate in-plane optical modes at the $\bm{\Gamma}$ point of the FBZ, and the second order peaks D+D'', 2D, and 2D', which are explained within the double-resonance Raman scheme. Being the sample pristine the defect-induced peaks D, D' and D'' are not visible, and indeed the double-resonance peaks are overtones of the defect peaks, as their nomenclature suggests (see Ref.~\cite{ferrari2013raman} for a historical overview of the understanding of the resonance Raman spectrum of graphene and the evolution of the nomenclature of the peaks).

\subsection{Double-resonance scattering intensity}
\subsubsection{Dependence of the linewidth on the trigonal warping}\label{subsec:trigWarp}
\begin{figure}[hbt!]
    \centering
    \includegraphics[width=1.\linewidth]{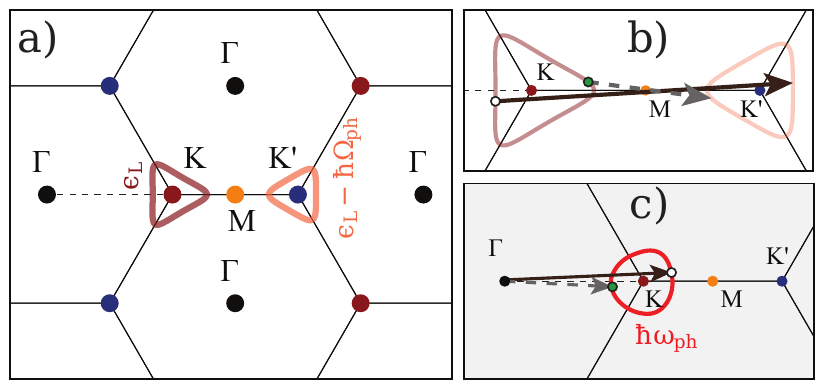}
    \caption{Illustration of the resonant scattering processes in the electronic (unshaded, panels a and b) and phononic (shaded, panel c) reciprocal spaces. \textbf{a)} Trigonal warping of the electronic bands shown by the iso-energy contours around the $\mathbf{K}$ and $\mathbf{K'}$ points, at $\epsilon_L$ and $\epsilon_L - \hbar\Omega^{\textrm{tot}}_\mathrm{ph}$, where $\hbar\Omega^{\textrm{tot}}_\mathrm{ph}$ is the total energy of the two scattered phonons, respectively. \textbf{b)} Zoom in on the relevant region of the reciprocal space. Two inter-band scattering processes (both contributing to the 2D peak) are shown: one where the initial electron-hole pair (white dot) is generated near the $\bm{\Gamma}$-\textbf{K} line (solid black arrow) and one where the pair (green dot) is generated near the \textbf{K}-\textbf{M} line (dashed grey arrow). \textbf{c)} In the case of perfect trigonal warping compensation, in the phononic reciprocal space the two scattering processes select phonons on the energy iso-contour $\hbar\omega_\mathrm{ph}$ having momenta close to \textbf{K}, either near the \textbf{K}-\textbf{M} line (white dot, solid black arrow, usually referred as \emph{outer}~\cite{berciaud2013intrinsic}), or the $\bm{\Gamma}$-\textbf{K} line (green dot, dashed grey arrow, referred as \emph{inner}). The two arrows are exactly the same as shown in panel b, but in the phononic reciprocal space their origin is $\bm{\Gamma}$.}
    \label{fig:trigwarp}
\end{figure}

As it was already discussed in Ref.~\cite{venezuela2011theory}, the striking narrow width of the double-resonance 2D and 2D' Raman lines is mainly attributed to an almost perfect compensation of the trigonal warping of the electronic and phononic dispersions and to a negligible effect of the electron-hole asymmetry for excitation energies $\epsilon_L<\SI{1.8}{eV}$. 
At a given laser excitation energy $\epsilon_L$, the compensation of trigonal warping occurs when the contour of the resonance phononic wavevectors coincides with an iso-energy contour of the phonon dispersion, thus narrowing the frequency distribution of the phonons involved.
The resonance wavevectors $\tilde{\vb{q}}$ are such that $I_{\nu\mu}(\tilde{\vb{q}})$ defined in Eq.~\ref{eq:Iq} is a maximum: given that the phonon dispersion plays a negligible role in the values of $I_{\nu\mu}(\tilde{\vb{q}})$ (so that Einstein phonons give almost the same results), the contour of $\tilde{\vb{q}}$ is entirely defined by the electronic dispersion, and in particular by its trigonal warping, as illustrated in Fig.~\ref{fig:trigwarp}.

\begin{figure}[hbt!]
    \centering
    \includegraphics[width=1.\linewidth]{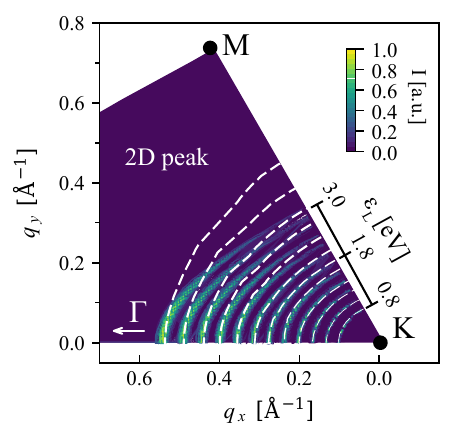}
    \caption{Intensity (for the 2D peak only) as a function of phonon wavevector $\vb{q}$ in the irreducible wedge of the FBZ, for different excitation energies between \SI{0.8}{eV} and \SI{3.0}{eV}, in steps of \SI{0.2}{eV}. The bright green region corresponds to the resonance wavevectors $\tilde{\vb{q}}$. Notice that it becomes more and more trigonally warped as we move away from the \textbf{K} point, i.e.\ for higher $\epsilon_L$. The white dashed lines indicate the phonon iso-energy contour: notice how for $\epsilon_L > \SI{1.8}{eV}$ their shape deviates significantly from the resonance $\tilde{\vb{q}}$ region.}
    \label{fig:Iq}
\end{figure}

In Figure~\ref{fig:Iq} we display the Raman intensity as a function of phonon wavevector (reduced to the irreducible wedge of the FBZ) for different excitation energies between \SI{0.8}{eV} and \SI{3.0}{eV}, and for the 2D peak only, that is $\mathcal{I}(\vb{q}) = \sum'_{\nu,\mu} I_{\nu\mu}(\vb{q})$ where the summation is restricted to the energy window of the 2D line, compared to the phononic iso-energy contours. One can clearly see how the contour of the resonance wavevectors $\tilde{\vb{q}}$ becomes more and more distorted (due to the trigonal warping of the electronic dispersion) at increasing excitation energies. In the same figure the iso-energy contours of the phonon dispersion are shown as white dashed lines, and they also become more distorted at increasing excitation energies, even though in a different fashion with respect to $\tilde{\vb{q}}$. As it was already pointed out in Ref.~\cite{venezuela2011theory}, the largest contribution to the Raman intensity comes from phonons having momenta along the $\boldsymbol{\Gamma}$-\textbf{K} line (which in literature are usually referred to as \emph{inner} phonons~\cite{berciaud2013intrinsic}, as opposed to \emph{outer} phonons, which have momenta along the \textbf{K}-\textbf{M} line), although one truly has to take into account the whole resonance region to properly describe the width of the peak. 

\begin{figure}[hbt!]
    \centering
    \includegraphics[width=0.8\linewidth]{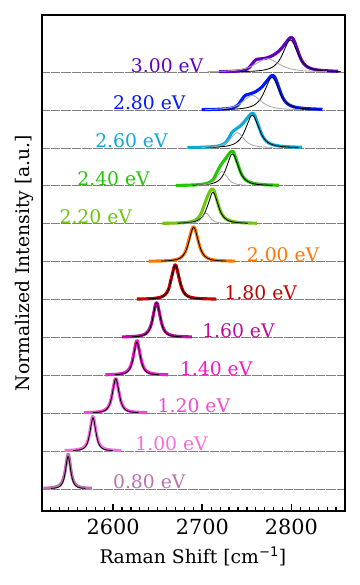}
    \caption{Shape of the 2D peak. All the peaks have been normalized to the height of their maxima. The black curves are Baskovian fitting functions, and we have separated the shoulder contribution for $\epsilon_L \geq \SI{2.2}{eV}$ (grey curves).}
    \label{fig:2Dpeak_RamanSpectrum}
\end{figure}

\begin{figure}[hbt!]
    \centering
    \hspace{-1cm}
    \includegraphics[width=0.9\linewidth]{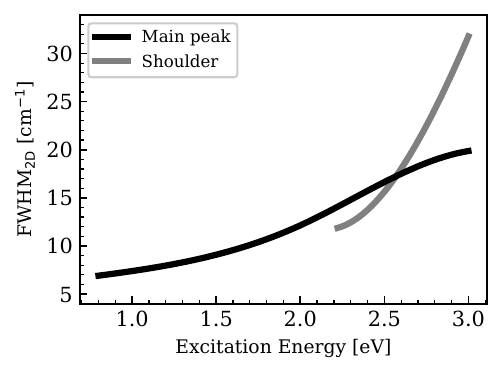}
    \caption{FWHM of the 2D peak as a function of the laser excitation energy, as extracted via the fitting with Baskovian functions (black line). For $\epsilon_L \geq \SI{2.2}{eV}$ we have separated the contribution to the FWHM of the shoulder at lower energies, which would give a much higher value of the width (grey line). We remark that the procedure of fitting the 2D peak via the sum of two Baskovian functions is customary in analysing experimental data~\cite{berciaud2013intrinsic} as well.}
    \label{fig:FWHM_2D}
\end{figure}

Indeed, the details of the trigonal warping of the phonon dispersion bear a strong influence on the FWHM of the 2D peak, in the sense that if, at fixed $\epsilon_L$, the iso-energy contours of the phonon dispersion do not lie entirely in the wavevector resonance region, then the resonance condition will select phonons having different frequencies, thus broadening the 2D peak. Notice that moving closer to the \textbf{K} point (that is lowering $\epsilon_L$) both the electronic and the phononic dispersions become more conically symmetric, since the asymmetry term scales as $(\vb{q}-\mathrm{\textbf{K}})^2$ \cite{gruneis2009phonon,basko2008theory}, and the compensation is guaranteed. On the other hand, for higher excitation energies the compensation becomes much worse, and eventually in the 2D peak a low-energy `shoulder' develops for $\epsilon_L \gtrsim \SI{2.2}{eV}$ (see Fig.~\ref{fig:2Dpeak_RamanSpectrum}). In Figure~\ref{fig:FWHM_2D} we report the FWHM of the 2D `main peak' and of its shoulder as a function of the excitation energy, as obtained via a fit with the sum of two Baskovian functions: it is evident that for larger $\epsilon_L$ the FWHM increases, as a consequence of the worse trigonal warping compensation. Indeed, the choice of fitting the 2D peak via the sum of two Baskovian functions is also customary when discussing experimental results~\cite{berciaud2013intrinsic}. In the following and if not specified otherwise, we take the FWHM as the one of the main peak.

\begin{figure}[hbt!]
    \centering
    \hspace{-1cm}
    \includegraphics[width=0.9\linewidth]{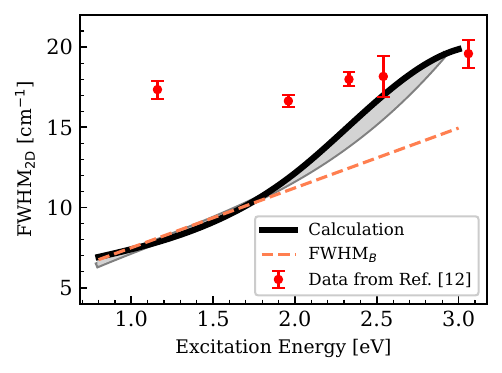}
    \caption{FWHM of the 2D peak as a function of the laser excitation energy, with perfect compensation of the trigonal warping, as obtained directly from the peak (gray solid line). The grey area indicates how the result may change for inexact compensation of the trigonal warping, and the black solid line is the result of the full calculation displayed in Fig.~\ref{fig:FWHM_2D} (`main peak'). The red squares indicate the measurements on hBN-encapsulated monolayer graphene~\cite{graziotto2022probing}. The orange dashed line is the result of Eq.~\ref{eq:FWHM_Baskovian} with $\gamma = \gamma_\mathrm{tot}$, shifted up by the inverse lifetime of the scattered phonon pair (i.e.\ \SI{5.3}{cm^{-1}}). Notice how for $\epsilon_L < \SI{1.8}{eV}$ Eq.~\ref{eq:FWHM_Baskovian} reproduces the result obtained for perfect trigonal warping compensation, as indeed should be the case since it has been obtained in the conical model approximation, where no trigonal warping is present. For $\epsilon_L > \SI{1.8}{eV}$, on the other hand, the electron-hole asymmetry plays the major role in increasing the width of the peak.}
    \label{fig:FWHM_compensated}
\end{figure}

Since the trigonal warping compensation depends mainly on the actual shape of the phonon dispersion, which is known theoretically up to a certain approximation (GW corrections to DFT as described in Ref.~\cite{herziger2014two}), we can provide a lower bound for the FWHM of the 2D peak by \emph{imposing} the perfect compensation of the trigonal warping by a geometrical argument. We proceed as follows: since the position of the resonance phonon wavevector is independent of the actual shape of the phonon dispersion, we are free to choose a phonon dispersion whose shape is suited to compensate the trigonal warping. The simplest analytical expression for the trigonally warped dispersion is 
\begin{equation}\label{eq:analyticalPhononDispersion}
    \hbar \omega_{\vb{q}} = P_0 + P\abs{\vb{q}} - (A - B \cos(3\phi_{\vb{q}})) \abs{\vb{q}}^2,
\end{equation}
where $\phi_{\vb{q}}$ is the angle formed by $\vb{q}$ with the $\boldsymbol{\Gamma}$-\textbf{K}-\textbf{M} line, and $P_0, P, A, B$ are excitation energy-dependent parameters to be obtained. Notice that imposing the compensation of the trigonal warping fixes only the values of the parameters $A$ and $B$ (that is, the shape of the iso-energy contours), since the values of $P_0$ and $P$ do not influence the asymmetry of the phonon dispersion, and can be chosen at will (e.g.\ from the experimentally determined position of the 2D peak). We can determine the parameters $A$ and $B$ by fitting Eq.~\ref{eq:analyticalPhononDispersion} to the resonance contour displayed as a bright green region in Fig.~\ref{fig:Iq}, obtaining $A = \SI{17.1}{cm^{-1} \text{\AA}^2}$, $B = \SI{120.7}{cm^{-1}\text{\AA}^2}$.\\
In Fig.~\ref{fig:FWHM_compensated} we report the 2D peak FWHM, as directly extracted from the peak, for the case where the phonon dispersion is imposed to best match the resonance contour of phonon wavevectors. In this case the 2D peak does not develop any shoulder at all laser energies. However, the FWHM of the 2D main peak computed with the GW corrected phonons, even if the trigonal warping matching is not perfect, has a very similar behaviour. This shows the importance of employing GW corrections in order to obtain very narrow peaks that match the experimental one in the visible region. In fact, in the same Figure we also display the FWHM of the 2D peak as obtained from measurements on hBN-encapsulated monolayer graphene~\cite{graziotto2022probing}, and it is immediately clear that while our model works in the visible, it fails to predict that the width of the peak does not decrease for lower excitation energies, but rather it stays almost constant between $16\text{-}18$ \SI{}{cm^{-1}}. As we will see in the following, this inconsistency will be solved when considering an enhancement of the EPC at lower excitation energies.

\begin{figure} [hbt!]
    \centering
    \hspace{-1cm}
    \includegraphics[width=0.9\linewidth]{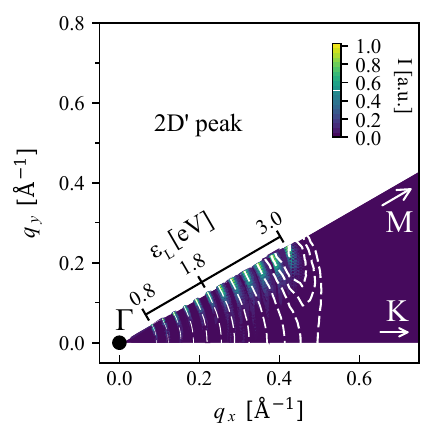}
    \caption{Intensity (for the 2D' peak only) as a function of phonon wavevector $\vb{q}$ in the irreducible wedge of the FBZ, for different excitation energies between \SI{0.8}{eV} and \SI{3.0}{eV}, in steps of \SI{0.2}{eV}. The bright green region corresponds to the resonance wavevectors $\tilde{\vb{q}}$: notice that the resonance wavevectors are basically along the $\boldsymbol{\Gamma}$-\textbf{M} line. The white dashed lines indicate the phonon iso-energy contour: close to $\epsilon_L = \SI{3.0}{eV}$ the scattered phonon comes from the highest point of the dispersion, hence the iso-contours form closed curves.}
    \label{fig:Iq_2Dpr}
\end{figure}

It is worth mentioning that for the 2D' peak we find instead a constant behaviour of the FWHM as a function of the excitation energy (the width stays between $6\text{-}7$ \SI{}{cm^{-1}} for $\epsilon_L$ between $0.8\text{-}3.0$ \SI{}{eV}), in agreement with the experimental data~\cite{graziotto2022probing}. Although near the $\bm{\Gamma}$ point in the FBZ the phonon dispersion is circularly symmetric, due to the fact that the Kohn anomaly is much weaker with respect to the anomaly at \textbf{K}, the main reason of the 2D' peak narrowness is that the double-resonance condition selects almost only phonons with wavevectors lying along the $\boldsymbol{\Gamma}$-\textbf{M} line (as already noticed in Ref.~\cite{venezuela2011theory}, this is partly due to the role of the matrix elements, and partly due to the fact that the resonance phonon wavevector needs to connect electronic states placed on opposite sides of the same trigonally warped Dirac cone), thus there is no need for the phonon dispersion to match the contour of the resonance wavevectors (see Figure~\ref{fig:Iq_2Dpr}, where we plot $\mathcal{I}(\vb{q})$, choosing the energy window corresponding to the 2D' peak). 

\subsubsection{Dependence of the linewidth on the electron-hole asymmetry}
Eq.~\ref{eq:Baskovian} (or \ref{eq:Baskovianform}) is obtained by considering conical electronic bands and neglecting the broadening of the line due to the finite lifetime of the phonon states. Including the latter means to convolve the Baskovian with the Lorentzian which replaces the $\delta$ function in Eq.~\ref{eq:RamanInt}, and to a first approximation one may consider the total FWHM as the sum of $\mathrm{FWHM}_B$ and the width of the Lorentzian. In our calculation we have verified that both the 2D peak (up to $\epsilon_L \simeq \SI{2.0}{eV}$, for higher $\epsilon_L$ a shoulder appears, see Section~\ref{subsec:trigWarp}) and the 2D' peak (up to $\epsilon_L \simeq \SI{3.0}{eV}$) can be almost perfectly fitted by a single Baskovian line-shape. By varying the electronic inverse lifetime $\gamma_\mathrm{tot}$ from the value given by Eq.~\ref{eq:gamma_tot} to twice this value we have verified that, at small excitation energies $\epsilon_L < \SI{1.8}{eV}$, the FWHM of both the 2D and 2D' peaks depends linearly on $\gamma_\mathrm{tot}$, as predicted by Eq.~\ref{eq:FWHM_Baskovian}, once one takes into account a non-zero intercept due to the Lorentzian broadening of phonon states. For larger excitation energies we still obtain a linear behaviour, but the slope is no more related simply to the ratio $v_\mathrm{ph}/v_F$: one has indeed to take into account the electron-hole asymmetry, which leads to an overall broadening of the peak. In particular, by studying the intensity as a function of the phonon wavevector $\vb{q}$, e.g.\ for the 2D peak (Fig.~\ref{fig:Iq}), we notice a much broader peak along the $\bm{\Gamma}$-\textbf{K} direction compared to the \textbf{K}-\textbf{M} direction (for $\epsilon_L \gtrsim \SI{2.0}{eV}$): this is due to the fact that phonons having $\vb{q}$ along $\bm{\Gamma}$-\textbf{K} are scattered by electron-hole pairs having wavevector $\vb{k}$ along the \textbf{K}-\textbf{M} direction (see Fig.~\ref{fig:trigwarp}), and in this direction the electron-hole asymmetry is higher, as visible in Fig.~\ref{fig:TB_DFT_electrons}. For the 2D' peak we notice the same broadening of the $\mathcal{I}(\vb{q})$ (Fig.~\ref{fig:Iq_2Dpr}), but since the slope of the Kohn anomaly at $\bm{\Gamma}$ is much lower than the slope of the anomaly at \textbf{K} this broadening does not reflect substantially in a wider $I(\omega)$ line. 

\subsubsection{Integrated area as a function of the electronic lifetime}\label{subsubsec:A_gamma}
\begin{figure}[hbt!]
    \centering
    \includegraphics[width=\linewidth]{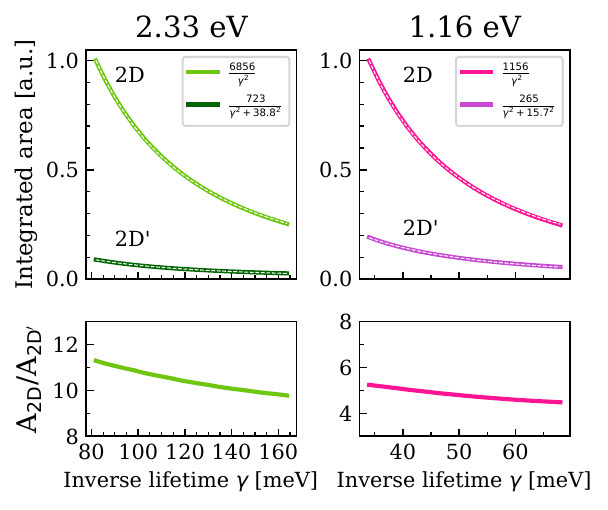}
    \caption{Top panels: Integrated area under the 2D and 2D' peaks for laser excitation energy of \SI{2.33}{eV} (left) and \SI{1.16}{eV} (right), as a function of the total inverse lifetime of the intermediate state $\gamma$. For fixed excitation energy both areas have been normalized to the area of the 2D peak at the lowest $\gamma$. The fitting functions are displayed in the legend (and shown as the perfectly superimposed white dashed lines): notice that while A$_\mathrm{2D}$ follows Ref.~\cite{basko2008theory}'s prediction of a decrease as $1/\gamma^2$, A$_\mathrm{2D'}$ follows this behaviour only approximately. The integration of the peak is performed on a range of ten times the FWHM centered on the maximum. Bottom panels: Ratio between the integrated areas under the 2D and 2D' peaks for laser excitation energy of \SI{2.33}{eV} (left) and \SI{1.16}{eV} (right), as a function of the total inverse lifetime $\gamma$.}
    \label{fig:areas_gamma}
\end{figure}

Eq.~\ref{eq:Baskovianform} predicts that, at fixed excitation energy $\epsilon_L$, the integrated area under the 2D and 2D' peaks, A$_\mathrm{2D}$ and A$_\mathrm{2D'}$, depends on the total inverse lifetime of the state $\gamma$ as $\text{A} = A_0/\gamma^2$, where $A_0$ is a constant~\cite{basko2008theory}. We have verified that in our calculation A$_\mathrm{2D}$ follows almost exactly the same behaviour (we integrate both the main peak and the shoulder), while A$_\mathrm{2D'}$ follows it only approximately. Indeed, we have calculated the Raman intensity by varying the value of $\gamma$ as a parameter, between the value given by Eq.~\ref{eq:gamma_tot} to twice this value (as we did in the previous section), and we display in Figure~\ref{fig:areas_gamma} (top panels) A$_\mathrm{2D}$ and A$_\mathrm{2D'}$ as a function of $\gamma$, for $\epsilon_L = \SI{1.16}{eV}$ and $\epsilon_L = \SI{2.33}{eV}$. We have performed a fitting via the functional form $A_0/(\gamma^2 + B^2)$, and we report the values of the $A_0$ and $B$ parameters in the legends of Fig.~\ref{fig:areas_gamma} (top panels). The discrepancy of the behaviour of A$_\mathrm{2D'}$ from the analytical prediction is most probably due to the fact that taking into account non-constant matrix elements over the FBZ results in a strongly anisotropic $I(\vb{q})$ near the $\bm{\Gamma}$ point (see Fig.~\ref{fig:Iq_2Dpr}). Despite the non-perfect adherence to the model predictions, our result improves on the conclusions reported in Ref.~\cite{venezuela2011theory}, thanks to the use of finer electronic and phononic wavevector grids.

However, looking at Figure~\ref{fig:areas_gamma} (bottom panels), where the ratio between the integrated areas under the 2D and 2D' peaks is reported as a function of $\gamma$, one can conclude that the ratio depends only weakly on $\gamma$ (within a $10\%$ error). As we will see below, this means that the experimental strong increase in the ratio $\text{A}_\mathrm{2D} / \text{A}_\mathrm{2D'}$ with lower excitation energies may be entirely attributed to the enhancement of the EPC.

\subsubsection{Integrated area as a function of the excitation energy}
Having confirmed that, at fixed excitation energy, the ratio between A$_\mathrm{2D}$ and A$_\mathrm{2D'}$ depends only slightly on the total inverse lifetime (see Section~\ref{subsubsec:A_gamma}) of the intermediate states, we can compare the result of our numerical calculation with the analytical result of Ref.~\cite{basko2008theory}, which predicts (notice that therein $\lambda_{\bm{\Gamma},\mathrm{\textbf{K}}} = \expval{g^2_{\bm{\Gamma},\mathrm{\textbf{K}}}}$ is used)
\begin{equation}\label{eq:basko_Aratio}
    \frac{\mathrm{A}_\mathrm{2D}}{\mathrm{A}_\mathrm{2D'}} (\omega_L) = 2 \left( \frac{\expval{g^2_{\mathrm{\textbf{K}}}}}{\expval{g^2_{\bm{\Gamma}}}} \frac{\omega_L - 2 \omega_\mathrm{\textbf{K}}}{\omega_L - 2\omega_{\bm{\Gamma}}}  \right)^2,
\end{equation}
where $\omega_{\mathrm{\textbf{K}}, \bm{\Gamma}}$ are the frequencies of the scattered phonons at the particular $\omega_L$, and the ratio $\expval{g^2_{\mathrm{\textbf{K}}}} / \expval{g^2_{\bm{\Gamma}}}$ is given in Ref.~\cite{basko2008theory} to be $1.14$, independent of the excitation energy.

\begin{figure}[hbt!]
    \centering
    \includegraphics[width=0.85\linewidth]{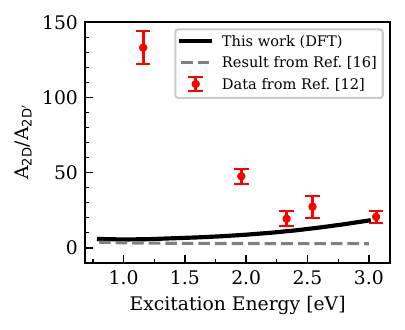}
    \caption{Ratio between the integrated areas under the 2D and 2D' peaks as a function of the excitation energy, as calculated with DFT ingredients in this work (apart from the correction to the electronic and phononic dispersions, and with inverse electron/hole lifetime given in Eq.~\ref{eq:gamma_Pedro}, black solid curve), and as given by the analytical result of Ref.~\cite{basko2008theory} (grey dashed curve). The red points indicate experimental data from Ref.~\cite{graziotto2022probing}, which are not in agreement with either calculation, in particular for excitation energies below \SI{2.3}{eV}.}
    \label{fig:AreaRatio_DFTonly} 
\end{figure}

In Figure~\ref{fig:AreaRatio_DFTonly} we compare the result of our calculation, i.e.\
\begin{equation}
    \frac{\mathrm{A}_\mathrm{2D}}{\mathrm{A}_\mathrm{2D'}} (\omega_L) = \frac{\int_\mathrm{2D} \omega^2 I(\omega) d\omega}{\int_\mathrm{2D'} \omega^2 I(\omega) d\omega},
\end{equation}
where $I(\omega) d\omega$ is the Raman scattering intensity as a function of the frequency of the scattered photon $\omega$ defined in Eq.~\ref{eq:IntensityFermi} and calculated with DFT only ingredients (except for the electronic and phononic dispersions, which have been corrected via the procedure explained in Section~\ref{sec:Comp_Appr}) in this work (the $\omega^2$ in the factor given by the photonic density of states, see Eq.~\ref{eq:intensity_withprefactors}), with the analytical result of Eq.~\ref{eq:basko_Aratio}. We notice that while Eq.~\ref{eq:basko_Aratio} predicts an almost constant behaviour as a function of the excitation energy, our numerical calculation shows an increase of the ratio with larger excitation energies, in particular for $\epsilon_L > \SI{1.5}{eV}$, which we mainly attribute to the role of the electron-hole asymmetry, which is neglected in the conical model of Ref.~\cite{basko2008theory}. The dependency of the ratio on the excitation energy was already discussed in Ref.~\cite{venezuela2011theory}, and in this work we have confirmed the behaviour also for excitation energies down to the infrared region, and employing finer electronic and phononic wavevectors grids. However, both calculations fail to reproduce the experimental data from Ref.~\cite{graziotto2022probing} (which are shown in Fig.~\ref{fig:AreaRatio_DFTonly} as red points), in particular for $\epsilon_L < \SI{2.3}{eV}$. From the analysis of the previous sections it is left as the only possible explanation an increase of the EPC with the lowering of the excitation energy~\cite{graziotto2022probing}, which we will discuss in the next section.

\subsubsection{EPC enhancement}
\begin{figure}[hbt!]
    \centering
    \includegraphics[width=0.85\linewidth]{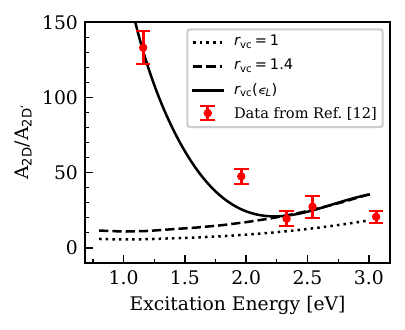}
    \caption{Ratio between the integrated areas under the 2D and 2D' peaks as a function of the excitation energy, as obtained by our calculation and multiplied by the $r_\mathrm{vc}^2$ rescaling factor. The black dotted curve is the same solid curve of Fig.~\ref{fig:AreaRatio_DFTonly} (since $r_\mathrm{vc} = 1$). The black dashed curve refers to the rescaling by the value obtained from GW calculations on graphite. The black solid curve indicates the rescaling by the excitation energy-dependent $r_\mathrm{vc}(\epsilon_L)$ defined in Eq.~\ref{eq:rVC_enhanced}.}
    \label{fig:AreaRatio_enhanced}
\end{figure}

As already anticipated in the previous section, the ratio between the integrated areas under the 2D and 2D' peaks gives crucial information about the EPC. In order to extract this information, in the previous section we have studied the behaviour of $\mathrm{A}_\mathrm{2D} / \mathrm{A}_\mathrm{2D'}$ as a function of the excitation energy, and found that we need to introduce a rescaling factor of the EPC in order to explain the strong increase of the ratio for lower excitation energies. A first attempt is to consider the EPC enhancement as given by GW calculations on graphite in Ref.~\cite{lazzeri2008impact}, that is we multiply the ratio as found by our DFT calculation by the square (since two phonons are being considered) of the rescaling factor $r_\mathrm{vc}$ defined in Eq.~\ref{eq:rvc}, which for graphite evaluates within GW to $1.4$ (see black dashed curve in Figure~\ref{fig:AreaRatio_enhanced}). This rescaling is not enough to match the experimental data of Ref.~\cite{graziotto2022probing}, hence we have therein introduced an excitation energy-dependent $r_\mathrm{vc}$, defined by
\begin{equation}\label{eq:rVC_enhanced}
    r_\mathrm{vc}(\epsilon_L) = \begin{cases}
                                1.4 & \epsilon_L > \SI{2.33}{eV} \\
                                2.5(\epsilon_L - 2.33)^2 + 1.4 & \SI{1.1}{eV} < \epsilon_L < \SI{2.33}{eV},
                                \end{cases}
\end{equation}
which evaluates to $4.8$ for $\epsilon_L = \SI{1.16}{eV}$ (i.e.\ it matches the data point at the lowest excitation energy), and consists of a second-order polynomial for excitation energies up to \SI{2.33}{eV}. This functional form has been chosen since it is the simplest differentiable curve which interpolates between the data, but it does not rely on a theoretical understanding of the behaviour of the EPC enhancement, which will be the scope of a future work. In particular, by assuming that $\expval{D_{\bm{\Gamma}}^2} = \SI{58.6}{eV^2 / \text{\AA}^2}$ (the GW value obtained on graphite, which we employ in place of graphene's one since its calculation has been studied in more detail in Ref.~\cite{lazzeri2008impact}) is not affected by the EPC rescaling, we obtain the enhanced value of $\expval{D_\mathrm{\textbf{K}}^2} = \SI{562.6}{eV^2 / \text{\AA}^2}$.

\begin{figure}[hbt!]
    \centering
    \includegraphics[width=0.85\linewidth]{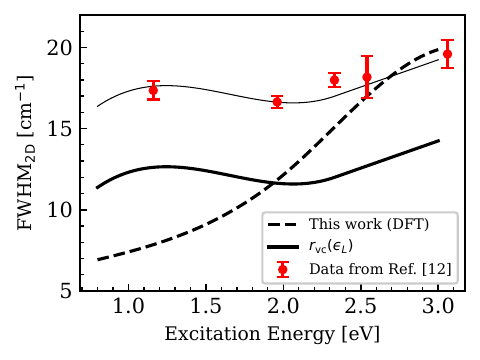}
    \caption{FWHM of the 2D peak as calculated in this work using the inverse electron/hole lifetime given by Eq.~\ref{eq:gamma_Pedro} (black dashed line, which is the same black solid curve in Fig.~\ref{fig:FWHM_compensated}), and the FWHM obtained employing the inverse lifetime given by Eq.~\ref{eq:gamma_enhanced}, which takes into account the enhancement of the EPC for lower excitation energies (black solid curve). Shifting up the latter curve by \SI{5}{cm^{-1}} (thin solid line) reproduces the behaviour of the experimental data of Ref.~\cite{graziotto2022probing} as a function of excitation energy.}
    \label{fig:FWHM_enhanced}
\end{figure}

We have then proceeded to calculate the effect of the EPC enhancement on the electron/hole inverse lifetime~\cite{graziotto2022probing}, as defined in Sec.~\ref{subsubsec:elec_hole_gamma}. As already evident from Fig.~\ref{fig:FWHM_compensated} our calculation is indeed not able to reproduce the almost constant behaviour of the FWHM of the experimentally measured 2D peak. Our previous analysis has excluded any role of the trigonal warping or of the electron-hole asymmetry in this failure, hence the major role must be played by the total inverse lifetime of the intermediate states. Assuming that the enhancement affects mainly the EPC near the \textbf{K} point, that is taking the value $\expval{g^2_{\bm{\Gamma}}} = \SI{0.052}{eV^2}$ from GW calculations on graphite, we can write
\begin{equation}
   \expval{g^2_\mathrm{\textbf{K}}} = 2 \expval{g^2_{\bm{\Gamma}}} \frac{\omega_{\bm{\Gamma}}}{\omega_\mathrm{\textbf{K}}} r_\mathrm{vc}.
\end{equation}
Substituting in Eq.~\ref{eq:elec_hole_gamma} we then obtain
\begin{equation}
\begin{split}
       \gamma^{\alpha\text{(e-ph)}}_\mathrm{enhanced} (\epsilon_L) = 0.248 \expval{g^2_{\bm{\Gamma}}} \Big[ \Big(1 + &\frac{\omega_{\bm{\Gamma}}}{\omega_\mathrm{\textbf{K}}} r_\mathrm{vc} \Big) \frac{\epsilon_L}{2} + \\
       & - \hbar\omega_{\bm{\Gamma}} (1+r_\mathrm{vc}) \Big] \SI{}{eV}  
\end{split}
\end{equation}
which using the values given in Tab.~\ref{tab:gamma_params} for the frequencies becomes
\begin{equation}\label{eq:gamma_enhanced}
       \gamma^{\alpha\text{(e-ph)}}_\mathrm{enhanced} = 13.2 \left[(1+1.30 r_\mathrm{vc})\frac{\epsilon_L}{2} - 0.196(1+r_\mathrm{vc}) \right] \SI{}{meV}.
\end{equation}
To model the impact of the inverse lifetime change on the FWHM we consider the expression for FWHM$_B$ obtained in Eq.~\ref{eq:FWHM_Baskovian}, which reproduces the result of our calculation for excitation energies below \SI{1.8}{eV} (see Fig.~\ref{fig:FWHM_compensated}), and we employ $\gamma = 2 \gamma^{\alpha\text{(e-ph)}}_\mathrm{enhanced}$, obtaining the black solid line in Figure~\ref{fig:FWHM_enhanced} which, shifted up by \SI{5}{cm^{-1}} (thinner solid line), clearly matches the almost constant behaviour of the experimental data as a function of the excitation energy. 

As a final remark one may inquire whether the EPC enhancement affects the slope of the Kohn anomaly, too. Indeed the real part of the phonon self-energy contributes to the energy shift $\Delta_{\vb{q}}^\nu$ of the harmonic phonon of mode $\nu$ with wavevector $\vb{q}$ via the following equation~\cite{calandra2005prb}
\begin{equation}
    \frac{\Delta_{\vb{q}}^\nu}{2} = \frac{2}{N_{\vb{k}}} \sum_{\vb{k},\alpha,\alpha'} \abs{g_{\vb{k}\alpha,\vb{k}+\vb{q}\alpha'}^\nu}^2 \mathcal{PV} \left[ \frac{f_{\vb{k}+\vb{q}\alpha'} - f_{\vb{k}\alpha}}{\epsilon_{\vb{k}+\vb{q}}^{\alpha'} - \epsilon^\alpha_{\vb{k}} -\omega_{\vb{q}}^\nu} \right],
\end{equation}
where $\mathcal{PV}$ is Cauchy's principal value, and $f$ is the Fermi distribution function. However, one can analytically compute the slope of the Kohn anomaly only under the assumption that the EPC is constant in the vicinity of the \textbf{K} point~\cite{piscanec2004kohn}, which as we have shown is not the case for graphene. Nonetheless, it can be argued that the enhancement of the EPC leads to a steepening of the Kohn anomaly, even though its quantification may be attained only accessing the full dependence of the EPC on the electronic wavevector.

\subsection{$r_{\mathrm{vc}}$ enhancement and resistivity}
As shown in Ref.~\cite{PhysRevB.90.125414}, the resistivity of graphene computed via the Boltzmann formalism using DFT or GW ingredients is underestimated in specific regimes, similarly to what happens for the ratio of the areas for the Raman spectrum. In particular, the experimental resistivity is significantly larger than the theoretical one especially at low dopings, but only for temperatures around room temperature and above, pointing to thermally-activated optical phonons as an underestimated scattering source.
Whether those optical phonons come from graphene itself or the substrate present in the transport measurement could not be definitely determined in Ref.~\cite{PhysRevB.90.125414}. However, it was argued that, since the coupling with substrate phonons is field-mediated, it would be strongly electrostatically screened for the carrier concentrations relevant in phonon-limited transport studies. Thus, the increase of resistivity was attributed to the increase of the intrinsic coupling between electrons and optical phonons of graphene at $\mathbf{K}$, namely $\beta^2_{K}$, beyond its GW value. The enhancement of $\beta^2_{K}$ was fitted on transport measurements, while the coupling with the zone-center modes ($\beta^2_O$) was not enhanced beyond its GW value. Since the couplings with optical modes involved in transport and Raman processes are related by the following relations~\cite{piscanec2004kohn,PhysRevB.90.125414}
\begin{align}
\beta^2_O=2\langle D^2_{\boldsymbol{\Gamma}}\rangle, \quad  \beta^2_K=\langle D^2_{\mathbf{K}} \rangle,
\label{eq:strD}
\end{align}
which allow to express the ratio of the scattering couplings in terms of $r_{vc}$, we can qualitatively support such conclusion in light of the present Raman data.

Besides the optical phonon frequency (which is the same in both Raman and transport setups), there are two important energy scales for electron-phonon interactions and $r_\mathrm{vc}$: the energy of the electron being scattered $\epsilon_{\vb{k}}$ and the Fermi level $\epsilon_F$ characterizing the doping.
In the transport case, those two energy scales coincide. Indeed, the electronic states participating to resistivity are within a relatively small energy window of order $k_B T$ around the Fermi level.
In the Raman data considered here, there is no doping~\cite{graziotto2022probing} such that the Fermi level can be set to zero, while the energy of the scattered electrons is half the laser energy, since the resonance condition for the electronic contour is expressed as $\epsilon_L=2\hbar v_F k$.
Considering these two energy scales, the mechanisms involving electron-phonon interactions can be compared in both experiments. In particular, the scattering processes relevant for resistivity satisfy the energy conservation of Eq.~56 of Ref.~\cite{PhysRevB.90.125414} restricted to intraband transitions.  
At the Raman resonance ($\epsilon_L=2\hbar v_F k$), the imaginary part of the second denominator of Eqs.~\ref{eq:Kaa} and \ref{eq:Kab}, in the limit of $\gamma\rightarrow0$, reduces to the same intraband delta function of Eq.~56 of Ref.~\cite{PhysRevB.90.125414} related to the emission of a phonon. 

\begin{figure}[hbt!]
    \centering
    \includegraphics[width=\linewidth]{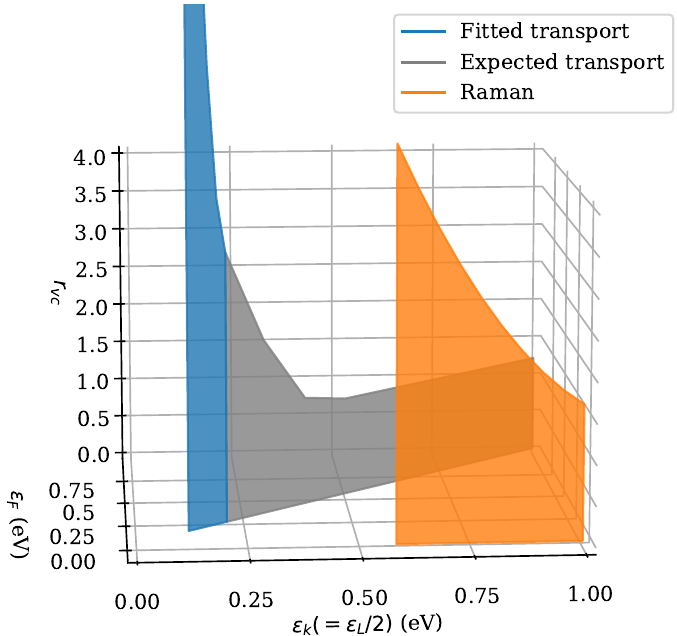}
    \caption{Electron-phonon coupling enhancement factor $r_\mathrm{vc}$ deduced from resistivity (blue) and Raman (orange) experiments. The grey curve represents the extrapolated expected behavior of the enhancement at large doping, i.e.\ returning to the \emph{ab initio} GW value.}
    \label{fig:res}
\end{figure}

We thus report in Fig.~\ref{fig:res} the fit of the couplings of Ref.~\cite{PhysRevB.90.125414} as a function of the doping level, translated in terms of the enhancement factor $r_{\mathrm{vc}}$. We also report the fit of $r_{\mathrm{vc}}$ on Raman measurements as discussed in this work. Note that the result of the enhancement of $r_{\mathrm{vc}}$ seen in Raman experiments is mostly independent on the type/presence of a substrate~\cite{graziotto2022probing}.

Fig.~\ref{fig:res} first shows that the electron-phonon coupling enhancements $r_\mathrm{vc}$ deduced from Raman and resistivity experiments are generally comparable, implying a potential explanation of the resistivity increase without resorting to extrinsic phonon contributions.
One may compare the transport- and Raman-deduced values of the enhancement at a fixed value of $\epsilon_{\vb{k}}$, keeping in mind that the Fermi levels $\epsilon_F$, and therefore the charge configurations of the system, are different ($\epsilon_F \approx \epsilon_{\vb{k}}$ for transport, $\epsilon_F \approx 0$ for Raman). In that case, the Raman-deduced value from this work is much larger than the resistivity one. As hinted in Ref.~\cite{basko2008interplay}, this is at least qualitatively expected. At large Fermi levels, the additional free-carriers are expected to strongly screen the enhancement, while at vanishing  Fermi levels this is not the case. This dependency on doping is further supported by the sharp increase of the enhancement in transport measurements close to the Dirac point.

\section{Conclusions and outlook}\label{sec:Conc}
In this work we have studied the double-resonance Raman intensity of monolayer graphene down to infrared laser energies via the use of first principles techniques. We found that both the trigonal warping of the electronic and phononic dispersions and the electron-hole asymmetry play a fundamental role in the determination of the line-shape and line-width of the 2D and 2D' peaks, and on their intensity as a function of the electron-hole lifetime. Keeping these effects in account, we are able to confidently justify the zone-boundary electron-phonon enhancement found in Ref.~\cite{graziotto2022probing} for laser energies in the infrared light spectrum. We have also addressed the consequences of such enhancement on the resistivity of graphene at room temperature, hinting towards a reconciliation of theoretical and experimental results. We hope that this work shall promote the interest in both performing Raman spectrum measurements nearer to the Dirac cone and in predicting the electron-phonon enhancement via the use of more refined theoretical many-body techniques: recent work shows indeed, by investigating the massless Dirac fermions of bilayer graphene, that the scale governing the enhancement is the vicinity in momentum to the Dirac point rather than the smallness of the electron-hole pair energy~\cite{graziotto2023enhanced}.

\section{Acknowledgments}
We thank Leonetta Baldassarre, Tommaso Venanzi, and Simone Sotgiu for the fruitful collaboration in the experimental investigation, and for beneficial discussions.
L.G.\ acknowledges funding from the Swiss National Science Foundation (SNF project no.\ 200020\_207795). We acknowledge the European Union's Horizon 2020 research and innovation program under grant agreements no.\ 881603-Graphene Core3 and the MORE-TEM ERC-SYN project, grant agreement no.\ 951215. We acknowledge PRACE for awarding us access to Joliot-Curie Rome at TGCC, France. Part of the calculations were performed on the DECI resource \emph{Mahti CSC} based in Finland at
\url{https://research.csc.fi/-/mahti}, and part on the ETH \emph{Euler} cluster at \url{https://scicomp.ethz.ch/}. Co-funded by the European Union (ERC, DELIGHT, 101052708). Views and
opinions expressed are however those of the authors only and do not
necessarily reflect those of the European Union or the European Research
Council. Neither the European Union nor the granting authority can be
held responsible for them.


\appendix
\section{Matrix element derivation}\label{app:matrixEl}
To obtain Eq.~\ref{eq:FermiGoldenRule} in the main text, we consider the operator $S$ that evolves the initial state into the final state, which is formally given~\cite{fetter2003quantum} by
\begin{equation}\label{eq:Sop}
\begin{split}
S = \sum_{n=0}^\infty \left(-\frac{i}{\hbar}\right)^n \int_{-\infty}^\infty dt_1 \int_{-\infty}^{t_1} dt_2 \cdots \int_{-\infty}^{t_{n-1}} dt_n \\ 
e^{-\varepsilon(\abs{t_1}+\dots+\abs{t_n})} \big(\mathcal{H}_I(t_1)\cdots\mathcal{H}_I(t_n)\big),
\end{split}
\end{equation}
with the limit $\varepsilon \to 0$ to be taken at the end. Notice in particular the absence of the $1/n!$ factor and the upper extrema of integration, which are different from infinity, since we are not employing the $\mathcal{T}$-product. As discussed before, we will limit ourselves to the fourth perturbative order in $\mathcal{H}_I(t)=e^{i\mathcal{H}_0t/\hbar}\mathcal{H}_I e^{-i\mathcal{H}_0t/\hbar}$, with $\mathcal{H}_I$ being either the electron-photon or the electron-phonon interaction Hamiltonian in the Schrödinger representation, and $\mathcal{H}_0 = \mathcal{H}_\mathrm{KS} + \mathcal{H}_\mathrm{em}$. We then obtain the matrix element given in Eq.~\ref{eq:FermiGoldenRule} in the main text by inserting complete sets of electron, phonon, and photon states, and integrating over the time variables.

\begin{figure}[hbt!]
    \centering
    \includegraphics[width=\linewidth]{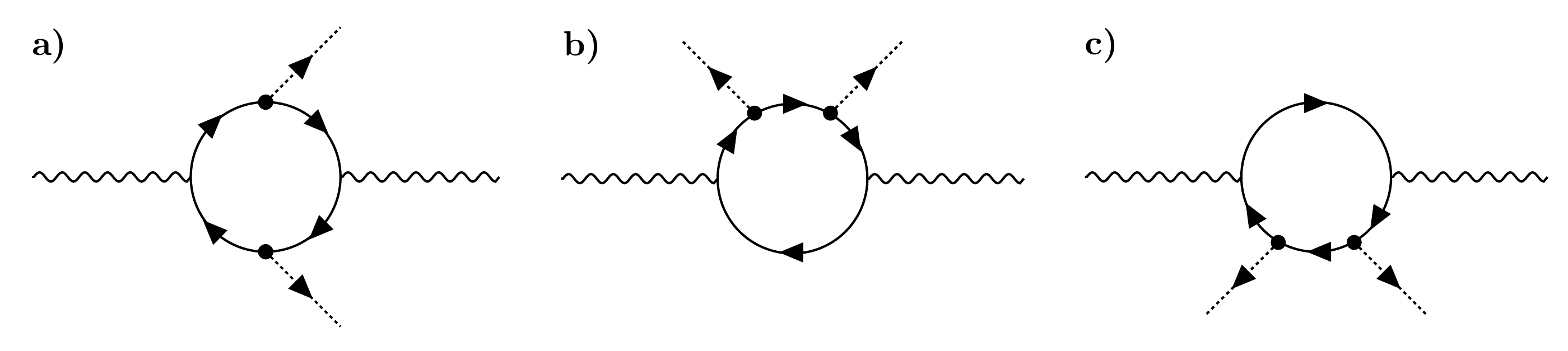}
    \caption{Feynman diagrams considered in the formalism of Ref.~\cite{basko2008theory}.}
    \label{fig:basko_diagrams}
\end{figure}

Notice that the choice of not employing the $\mathcal{T}$-product in Eq.~\ref{eq:Sop} (as opposed to the formalism of Ref.~\cite{basko2008theory}) means that we have to consider in Eq.~\ref{eq:FermiGoldenRule} all the permutations of both the electron-phonon and electron-photon interaction Hamiltonians which give rise to non-zero matrix elements. That is, one has to  consider the arbitrary time-ordering of two electron-phonon and two electron-photon interaction vertices, where, for example, one may have that the phonon is emitted before the photon is absorbed. One can identify 3 topologically-inequivalent diagrams (see Figure~\ref{fig:basko_diagrams}), so that the total number of Stokes diagrams is $4!\times3$ (those depicted in Fig.~1b in Ref.~\cite{PhysRevB.8.2795}). On the other hand, in the formalism of Ref.~\cite{basko2008theory} only the three topologically-inequivalent diagrams need to be considered, since the time-ordering is already been taken care of by the presence of the $\mathcal{T}$-product. We anyway choose not to employ it since it is easier to enforce the resonance condition having an explicit time ordering of the vertexes, as discussed in the main text.

On the other hand, in Ref.~\cite{basko2008theory} the resonance condition is enforced by throwing away the diagrams (b) and (c) of Fig.~\ref{fig:basko_diagrams} (which contains the resonant diagrams indicated as \emph{ee} or \emph{hh} in Fig.~\ref{fig:venezuela_diagrams}, that are shown to be negligible with respect to the \emph{eh} and \emph{he} ones), and by neglecting the off-resonant contributions to the diagram (a) (see Eq.~58 of Ref.~\cite{basko2008theory}, where the approximation consists in eliminating the non-resonant denominator).

Finally we notice that, in both formalisms, in principle all vertexes but one (at choice) contain electronic screening~\cite{calandra2010adiabatic}. We will consider all light vertexes as unscreened in our calculation, exploiting the fact that the electronic screening of the electron-light matrix element is negligible in graphene for in-plane light polarizations~\cite{binci2021first}.

\section{Telescopic grids}\label{app:telescopic}

\begin{figure}
    \centering
    \includegraphics[width=\linewidth]{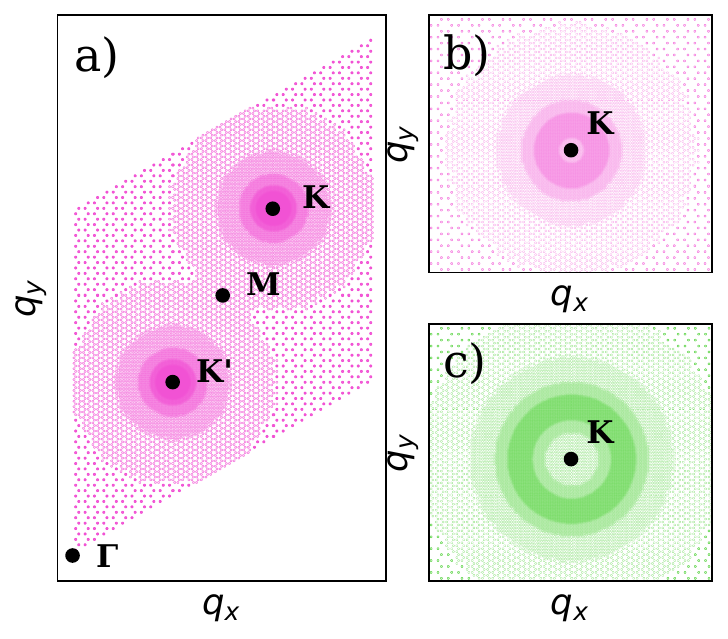}
    \caption{\textbf{a)} Telescopic grid which has been used to interpolate the electronic momenta for \SI{1.20}{eV} excitation energy. The parameters are $(l,L,\mathcal{N}) = (2,10,26)$. \textbf{b)} Zoom in on the resonance region around the \textbf{K} point. \textbf{c)} Zoom in on the resonance region of the grid employed for \SI{2.40}{eV} excitation energy. $(l, L, \mathcal{N}) = (2, 10, 23)$. The marker sizes are proportional to the weights of the momenta in the calculation.}
    \label{fig:Raman-grids}
\end{figure}

The grids on which the electronic momenta $\vb{k}$ have been Wannier-interpolated are built with a procedure analogous to the one described in Ref.~\cite{binci2021first}. Indeed an ultra-dense $\vb{k}$ mesh is desired to be located around the $\vb{k}$ resonance region, which lies in an annulus with average radius $\epsilon_L / (2 \hbar v_F)$, in order to achieve a fast convergence of the summation in Eq.~\ref{eq:Iq}. The procedure is the following: we generate the first $\vb{k}$ point (the \textbf{K} point) to be the center of an equilateral triangle with side $4\pi / (\sqrt{3} a)$, i.e.\ the modulus of the reciprocal lattice vectors (this defines the zeroth level, and the FBZ which we employ is just given by two adjacent equilateral triangle, see Figure~\ref{fig:Raman-grids}a).

 The first level is given by the four (including the one generated at the previous level) $\vb{k}$ points which are the baricenters of the four smaller equilateral triangles in which the zeroth order triangle is partitioned into (i.e.\ they are the three points located at the midpoint between the center of the zeroth order triangle and its three vertices, plus the center point itself). The procedure is iterated and at level $\ell$ the weight of the $\vb{k}$ point is given by $1/4^\ell$. One then defines a minimum depth $l$ and a maximum depth $L$, and the $\vb{k}$ points generated via the procedure above are kept only if their level is $ \leq l$, or if $l < \ell \leq L$ and at the same time the point lies in the resonance annulus within a range $(D / \mathcal{N}) 4^{(L-\ell)/p}$, where $D = 4\pi/(3a)$ is the distance from \textbf{K} to a vertex of the zeroth order triangle, $\mathcal{N}$ is an integer, and $p = 2$ has been chosen by making sure that the ultra-dense region fully includes the trigonally warped electronic iso-energy contour at the excitation energy. Therefore the grids are dependent on the excitation energy and are densified in the resonance annulus (so that in the resonance region they are as dense as a $1448 \times 1448$ uniform grid), at variance with the ones of Ref.~\cite{binci2021first}.

\bibliography{main}
\end{document}